\documentclass[pre,twocolumn,superscriptaddress,floatfix,showpacs,showkeys]{revtex4}

\usepackage[latin2]{inputenc}
\usepackage{amsmath}
\usepackage{amssymb}
\usepackage{epsfig}

\begin{document}

\title{Influence of polydispersity on micromechanics of granular materials}

\date{\today}

\author{M.\ Reza Shaebani}
\email{reza.shaebani@uni-due.de}
\affiliation{Department of Theoretical Physics, University of
Duisburg-Essen, 47048 Duisburg, Germany}

\author{Mahyar Madadi}
\email{mahyar.madadi@curtin.edu.au}
\affiliation{Department of Applied Mathematics, 
The Research School of Physics and Engineering, 
The Australian National University, Canberra 0200, Australia}
\affiliation{Department of Exploration Geophysics, Curtin 
University, P.O. Box U1987, Perth, WA 6845, Australia}

\author{Stefan Luding}
\email{s.luding@utwente.nl}
\affiliation{Multi Scale Mechanics (MSM), CTW, UTwente, 
P.O. Box 217, 7500 AE Enschede, Netherlands}

\author{Dietrich E.\ Wolf}
\affiliation{Department of Theoretical Physics, University of
Duisburg-Essen, 47048 Duisburg, Germany}

\begin{abstract}
We study the effect of polydispersity on the macroscopic
physical properties of granular packings in two and three
dimensions. A mean-field approach is developed to approximate 
the macroscale quantities as functions of the microscopic 
ones. We show that the trace of the fabric and stress tensors 
are proportional to the mean packing properties (e.g.\ packing
fraction, average coordination number, and average normal force)
and dimensionless correction factors, which depend only on the 
moments of the particle-size distribution. Similar results are 
obtained for the elements of the stiffness tensor of isotropic 
packings in the linear affine response regime. Our theoretical 
predictions are in good agreement with the simulation results.
\end{abstract}

%45.70.-n   Granular systems (see also 05.65.+b Self-organized systems)
%45.70.Cc   Static sandpiles; granular compaction
%83.80.Fg   Granular solids

\maketitle

\section{Introduction}
\label{Introduction}

The physics of granular media has received a lot of attention 
because of its scientific challenges and industrial relevance. 
The structural and dynamical properties of granular materials 
differ from those of ordinary solids, liquids, or gases due to 
nonlinearity and disorder \cite{deGennes99,Jaeger96,Hinrichsen04}. 
On the microscopic level, a static assembly of grains consists of 
particles which interact with their neighbors in order to prevent 
interpenetration. In spite of the uniform density of granular 
packings, the resulting contact and force networks between 
particles are highly inhomogeneous 
\cite{Radjai96,Makse00,Ostojic06}, leading to many intriguing 
phenomena in these systems. Describing the behavior via 
micromechanical approaches, in which the discrete nature of the 
system is taken into account, is thus commonly preferred to 
continuum-mechanical approaches where some heuristic assumptions 
have to be made in order to construct the constitutive equations 
for macroscopic fields. One can then express the macroscopic physical 
quantities in terms of the microscale ones. For example, thermal 
and electrical conductivities are related to the trace of the 
fabric tensor, a micro-geometrical probability of the 
orientations of contacts. While the relationship between 
macroscopic and microscopic properties of granular media has been 
studied widely \cite{deGennes99,Hinrichsen04,Luding04}, the 
question remains as to what extent the macroscale quantities 
are sensitive to the micro-scale details, and how large is the 
error introduced in the calculation of the ``observable 
quantities" by taking into account only the average packing 
properties.

Granular materials in nature and industry consist of particles
with the common property of polydispersity. It is known that size 
polydispersity affects the mechanical behavior of granular 
systems (e.g.\ shear strength) \cite{Wackenhut05,Voivret09} as 
well as their space-filling properties (e.g.\ packing fraction) 
\cite{Herrmann03,Voivret07}, which are crucial in many chemical 
processes like absorption, filtering, etc. Polydispersity in most 
studies, so far, has been restricted to narrow size distributions 
mainly to prevent long-range structural order; however, there are 
a few studies where broader ranges of particle-size distribution 
are investigated \cite{Voivret09,Voivret07,Dodds02,Ogarko12}. In 
this paper, we address the question of how deviation from the 
monodisperse case influences the macroscopic properties of 
granular assemblies.

We consider a special case of spherical particles [or disks in 
two dimensions (2D)] allowing for analytical calculations. The main 
goal is to develop a mean-field approach to calculate the 
desired microscopic quantities, such as the trace of the fabric 
and stress tensors, and the elements of the stiffness tensor 
in two- and three-dimensional polydisperse granular systems. 
These quantities are directly connected to macroscopic quantities 
such as thermal and electrical conductivities, isotropic pressure, 
and bulk and shear moduli. A similar analytical approach has been 
already used in Ref.~\cite{Madadi04} to calculate the trace of 
the fabric tensor in 2D packings, where it turned out that the 
trace of fabric is factorized into three contributions: (i) the 
volume fraction, (ii) the mean coordination number, and (iii) a
dimensionless correction factor which only depends on the particle-size 
distribution. Using a similar approach, here we investigate
also the stress and stiffness tensors and extend the method to 3D
cases. In order to compare the analytical results with numerical
simulations, we first construct static packings of grains using 
contact dynamics simulations \cite{Moreau94,Jean99,Brendel04}. 
The initial dilute systems of rigid particles are compressed by 
imposing a confining pressure to get the final static homogeneous 
packings \cite{Shaebani09}. Comparisons are then made 
between the results of our mean-field model and the exact values 
obtained from the numerical simulations.

This work is organized in the following manner: The fabric tensor
of a polydisperse assembly of spherical particles is investigated 
in Sec.~\ref{Fabric-Tensor}, and a mean-field approach is introduced
to calculate the trace of fabric. We present the analytical results 
for the calculation of the stress tensor in Sec.~\ref{Stress-Tensor}, 
and the same approach is used in Sec.~\ref{Stiffness-Tensor} to 
investigate the stiffness tensor elements in frictionless isotropic 
packings. In Sec.~\ref{Simulation-Results}, the analytical 
calculations are compared to numerical simulations of corresponding 
packings of polydisperse particles. Finally, we discuss and conclude 
the results in Sec.~\ref{Conclusion}. Detailed calculations for 
two-dimensional packings of disks are presented in the Appendix.

\section{Fabric tensor}
\label{Fabric-Tensor}

\subsection{Single-particle case}
\label{Fabric-Single}

Various definitions of the fabric tensor have been used in the 
literature to describe the spatial arrangement of the particles 
in a granular assembly \cite{Goddard86,Chang88,Rothenburg81}. 
The fabric tensor of the second order for one particle is defined as
\cite{Mehrabadi82,Latzel00,Madadi04}
\begin{equation}
 {\mathit h}^{^p}_{_{\alpha \beta}} = \sum_{c=1}^{C_p}
 \frac{l^{^{pc}}_\alpha}{|\vec l^{^{pc}}|} \frac{l^{^{pc}}_\beta}
 {|\vec l^{^{pc}}|},
  % \null_{\text{fabric-single}}
  \label{fabric-single}
\end{equation}
where $C_p$ is the number of contacts of particle $p$, and
$l^{^{pc}}_\alpha$ is the $\alpha$ component of the branch vector
$\vec l^{^{\; pc}}$, connecting the center of particle $p$ to its
contact $c$. In the case of spherical particles, the unit branch
vector $\vec l^{^{\; pc}} / |\vec l^{^{\; pc}}|$ and the unit
normal vector $\hat n^{^{pc}}$ at contact $c$ are identical. The
trace of the single-particle fabric tensor in a $D$-dimensional 
system is
\begin{equation}
 {\mathit h}^{^p}_{_{\alpha \alpha}} = \sum_{c=1}^{C_p} 
 \sum_{\alpha=1}^{D} \frac{l^{^{pc}}_\alpha}{|\vec l^{^{pc}}|} 
 \frac{l^{^{pc}}_\alpha} {|\vec l^{^{pc}}|} = C_p,
  % \null_{\text{fabric-single-trace}}
  \label{fabric-single-trace}
\end{equation}
i.e.\ the number of contacts of particle $p$.

\subsection{Many-particle case}
\label{Fabric-Many}

The average fabric tensor $\langle {\mathit h}_{_{\alpha \beta}}
\rangle_{_V}$ enables us to describe the global contact network
in a given volume $V$. Assuming that the contribution of particle 
$p$ (lying inside $V$) to the average fabric tensor is proportional 
to its volume $V_p$, we obtain
\begin{equation}
\langle {\mathit h}_{_{\alpha \beta}} \rangle_{_V} =
\frac{1}{V}\sum^N_{p=1} V_p {\mathit h}^{^p}_{_{\alpha \beta}},
  % \null_{\text{fabric-many}}
  \label{fabric-many}
\end{equation}
where the sum runs over all particles lying inside $V$, and
$\langle \cdot \cdot \cdot \rangle_{_V}$ denotes the volume
weighted average. Using Eq.~(\ref{fabric-single-trace}) to
calculate the trace of the average fabric tensor, we get
\begin{equation}
\langle {\mathit h}_{_{\alpha \alpha}} \rangle_{_V} =
\frac{1}{V}\sum^N_{p=1} V_p C_p,
  % \null_{\text{fabric-many-trace}}
  \label{fabric-many-trace}
\end{equation}
which can be interpreted as the contact number density.
Alternative possibilities, e.g.\ using the volume of the 
polygon that contains the particle (obtained e.g.\ via 
Voronoi tessellation), or introducing constant prefactors 
or slightly different volume contributions are not discussed 
here (see Refs.~\cite{Goddard86,Chang88,Cowin85,Duran10} for 
more details). In a monodisperse packing, 
Eq.~(\ref{fabric-many-trace}) for identical particles is 
reduced to $\langle {\mathit h}_{_{\alpha \alpha}} 
\rangle_{_V} \!=\! \phi z$, where $\phi$ is the packing
fraction ($\phi = \sum_{p} V_p / V$), and $z$ is the average
coordination number ($z = \sum_{p} C_p / N$). We note 
that only ``real" contacts contribute to the calculation 
of $z$, and geometrical neighbors without a permanent 
physical contact, which do not contribute in the fabric 
and force carrying structures, are not considered here. 

\subsection{Polydispersity}
\label{Fabric-Polydispersity}

\begin{figure}
\epsfig{figure=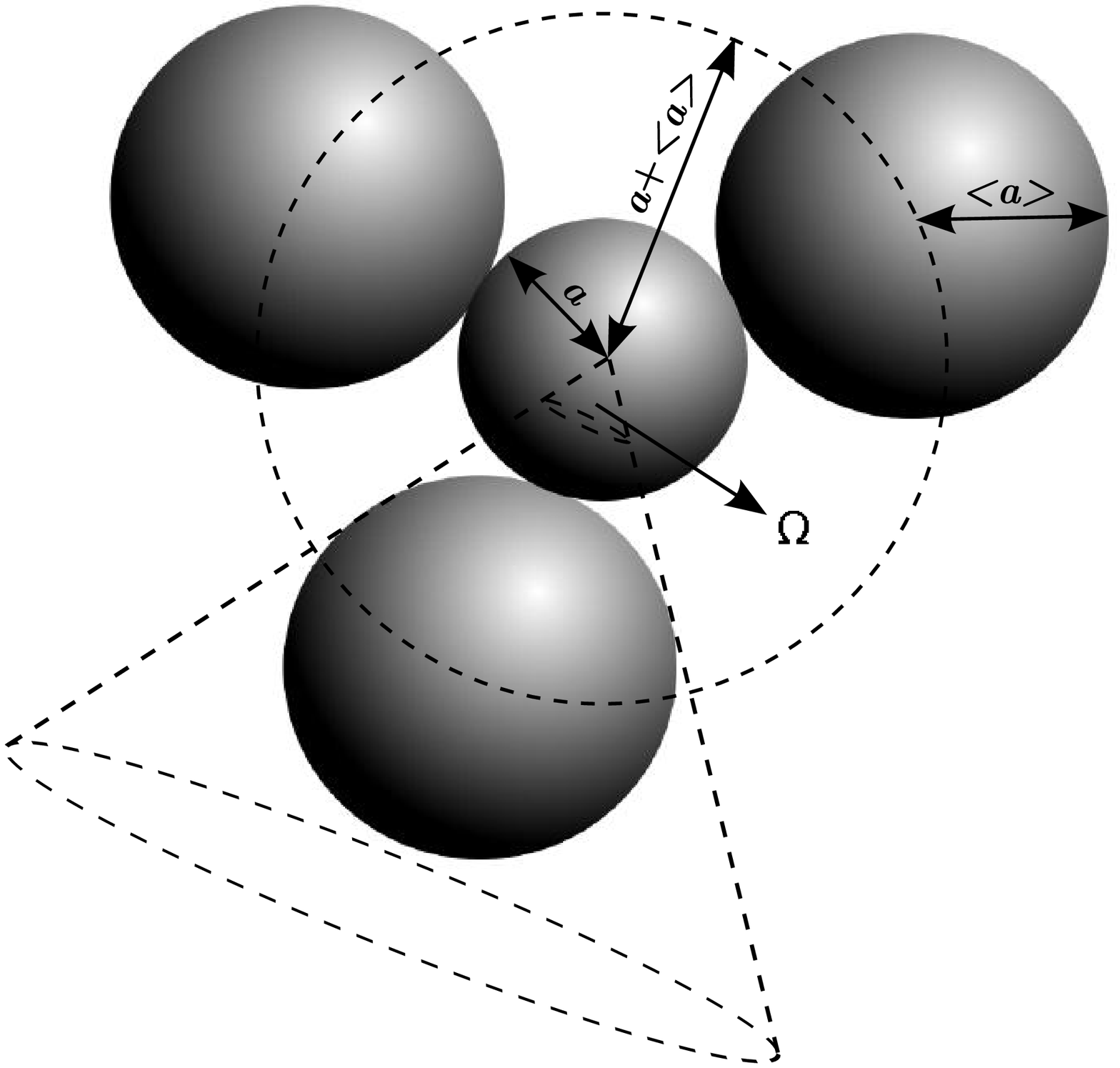,width=0.65\linewidth,angle=0}
\caption{Schematic picture showing a typical particle with radius
$a$ surrounded by identical particles of average radius $\langle
a \rangle$ in a 3D packing of spheres. 
\label{Fig-Schematic1}}
\end{figure}

For an accurate evaluation of the trace of the average fabric tensor 
in a polydisperse granular packing, one should take into account the 
contributions from all particles. However, if the distribution 
function of particle radii is known, $\langle {\mathit h}_{_{\alpha 
\alpha}} \rangle_{_V}$ can be approximated as a function of the 
moments of the size distribution. We assume a polydisperse 
distribution of particle radii with probability $f(a) da$ to find 
the radius between $a$ and $a+da$, and with $\int_0^\infty f(a)da=1$. 
The continuum limit of Eq.~(\ref{fabric-many-trace}) is then given by
\begin{equation}
\langle {\mathit h}_{_{\alpha \alpha}} \rangle_{_V} = \frac{N}{V}
\int_0^\infty V(a) C(a) f(a) da.
  % \null_{\text{fabric-poly-continuous}}
  \label{fabric-poly-continuous}
\end{equation}
Here, $C(a)$ is the average coordination number of particles with
radius $a$. We evaluate $C(a)$ using a mean-field approach
similar to the one proposed in \cite{Ouchiyama81} and used
already in \cite{Madadi04} to study the trace of the fabric
tensor. In the following, we concentrate on the case of spherical 
particles in three-dimensional systems (see the detailed calculations 
for two-dimensional packings of disks in the Appendix). Let us suppose 
that each particle in the polydisperse granular medium is surrounded 
by identical particles of average radius $\langle a \rangle$ (see 
Fig.~\ref{Fig-Schematic1}), where $\langle a \rangle \!=\! \int_{_0}^{^\infty} 
\!\!\! a f(a) da$.  The surface of a reference particle of radius $a$ 
is then shielded by its $C(a)$ neighboring particles of radius $\langle 
a \rangle$. The space angle covered by a neighboring particle on the 
reference particle in a three-dimensional packing of spheres is
\begin{equation}
\Omega(a) = 2 \pi \Bigg( 1-\frac{\sqrt{(a+\langle a \rangle)^2-
\langle a \rangle^2}}{a+\langle a \rangle} \Bigg).
  % \null_{\text{space-angle}}
  \label{space-angle}
\end{equation}
The total fraction of shielded surface, also called linear
compacity, is obtained as
\begin{equation}
c_s(a) = \frac{1}{4 \pi a^2} \sum_{i=1}^{C(a)}\Omega(a) 
a^2 = \Omega(a)C(a)/4 \pi.
% \null_{\text{fraction-shielded}}
\label{fraction-shielded}
\end{equation}
Now, another basic assumption is that the total fraction of 
shielded surface $c_s$ is independent of the particle radius 
$a$. As a result, the expected mean coordination number becomes
\begin{eqnarray}
z = \int_0^\infty \!\!\! C(a)f(a)da = 4 \pi c_s q_{_0},
% \null_{\text{mean-coordination-number}}
\label{mean-coordination-number}
\end{eqnarray}
with $ q_{_0} {=} \int_{_0}^{^\infty} \!\! f(a)/ \Omega(a) da$.
Using Eqs.~(\ref{fraction-shielded}) and
(\ref{mean-coordination-number}) one finds
\begin{equation}
C(a) = \frac{z}{q_{_0} \Omega(a)}.
  % \null_{\text{coordination-number}}
  \label{coordination-number}
\end{equation}
The trace of the fabric tensor for a polydisperse packing is then
obtained by substitution of Eq.~(\ref{coordination-number}) in
Eq.~(\ref{fabric-poly-continuous}),
\begin{equation}
\langle {\mathit h}_{_{\alpha \alpha}} \rangle_{_V} = \phi z g_{_1},
  % \null_{\text{fabric-meanfield}}
  \label{fabric-meanfield}
\end{equation}
where the correction factor $g_{_1}$ is defined as
\begin{eqnarray}
g_{_1} = \frac{\displaystyle\int_0^\infty V(a) \frac{f(a)}{\Omega(a)}
da}{q_{_0} \displaystyle\int_0^\infty V(a) f(a) da} = \frac{\langle 
a^3 \rangle_{_g}}{\langle a^3 \rangle}
% \null_{\text{fabric-correctfactor}}
\label{fabric-correctfactor}
\end{eqnarray}
Here, $\langle a^k \rangle \!$ and $\langle a^k \rangle_{_g}$ denote 
the $k$th moments of the size distribution $f(a)$ and the modified 
distribution $f(a)/\Omega(a)$ normalized by $q_{_0}$, respectively. 
We note that $g_{_1}$ depends only on the size distribution function 
$f(a)$.

\subsection{Narrow size distributions}
\label{Fabric-NarrowDistribution}

\begin{figure}
\centering
\includegraphics[scale=0.61,angle=0]{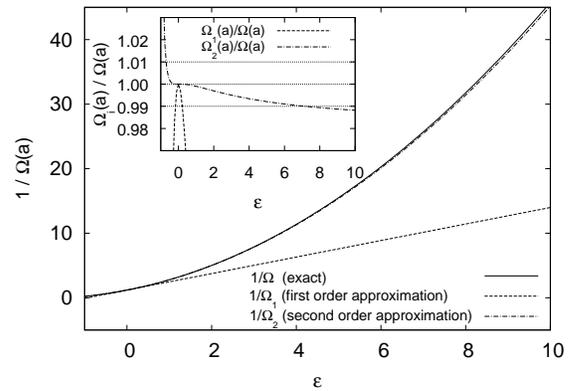}
\caption{$1/\Omega(a)$ as a function of $\epsilon$. The exact value 
(solid line) is compared with the first-order (dashed line) and 
second-order (dash-dotted line) approximations. The inset shows more clearly the 
deviation of the approximations from the exact value.}
\label{narrow-size-dist-1}
\end{figure}

By introducing $\epsilon(a) {=} a / \langle a \rangle{-}1$, which 
ranges between $-1$ and $\infty$ depending on the choice of $a$, 
Eq.~(\ref{space-angle}) can be written as
\begin{equation}
\Omega(a) = 2 \pi \Bigg(1-\frac{\sqrt{\epsilon^2+4\epsilon+3}}
{2+\epsilon} \Bigg).
  % \null_{\text{space-angle-epsilon}}
  \label{space-angle-epsilon}
\end{equation}
Indeed, $\epsilon(a)$ quantifies the deviation from the mean 
particle size $\langle a \rangle$ [e.g.\ $\epsilon(a)$ equals 
zero in the monodisperse case]. Hence, for narrow size distributions, 
we approximate $1/\Omega(a)$ by Taylor expansion around 
$\epsilon {=} 0$ (corresponding to Taylor expansion around 
$a {=} \langle a \rangle$). By Taylor expansion to the second 
order in $\epsilon$, one obtains
\begin{equation}
\frac{1}{\Omega(a)} \simeq A_{_1} + B_{_1}\epsilon +
C_{_1}\epsilon^2,
  % \null_{\text{space-angle-expansion}}
  \label{space-angle-expansion-3D}
\end{equation}
with $A_{_1} \!\!=\! \frac{1}{(2-\sqrt 3) \pi }$, $B_{_1} \!\!=\!
\frac{1}{2 \sqrt 3 (2-\sqrt 3)^2 \pi}$, and $C_{_1} \!\!=\!
\frac{1}{3(3+\sqrt 3)(2-\sqrt 3)^2\pi}$. The first-order
approximation deviates significantly from the exact value
(see Fig.~\ref{narrow-size-dist-1}). However, the second-order
expansion provides a good approximation with less than $1\%$
error in the range $-0.5\! < \!\epsilon\! < \!7.5$ (or $0.5
\langle a \rangle\! <\! a \!<\! 8.5 \langle a \rangle$).

Therefore, the correction factor
[Eq.~(\ref{fabric-correctfactor})] for narrow size distributions
becomes
\begin{equation}
g_{_1} \!\simeq\! \frac{(A_{_1} {-} B_{_1} {+} C_{_1}) {+} (B_{_1} 
{-} 2C_{_1}) \frac{\big\langle \displaystyle{a}^4 \big\rangle}
{\big\langle \displaystyle{a} \big\rangle \big\langle \displaystyle{a}^3 
\big\rangle} {+} C_{_1}\frac{\big\langle \displaystyle{a}^5 \big\rangle}
{\big\langle \displaystyle{a} \big\rangle^2 \big\langle \displaystyle{a}^3 
\big\rangle}}{(A_{_1} {-} C_{_1}){+} C_{_1}\frac{\big\langle \displaystyle{a}^2 
\big\rangle}{\big\langle \displaystyle{a} \big\rangle^2}}.
% \null_{\text{fabric-correctfactor-narrow}}
\label{fabric-correctfactor-narrow}
\end{equation}
Equation~(\ref{fabric-correctfactor-narrow}) should account for
arbitrarily shaped size distributions $f(a)$ as long as they are 
not too wide. Note the different nomenclature in Ref.\ \cite{Goncu10}, 
where the above equation is introduced with different abbreviations and 
coefficients.

\section{Stress tensor}
\label{Stress-Tensor}

\subsection{Single-particle case}
\label{Stress-Single}

\begin{figure}
\centering
\includegraphics[scale=0.34,angle=0]{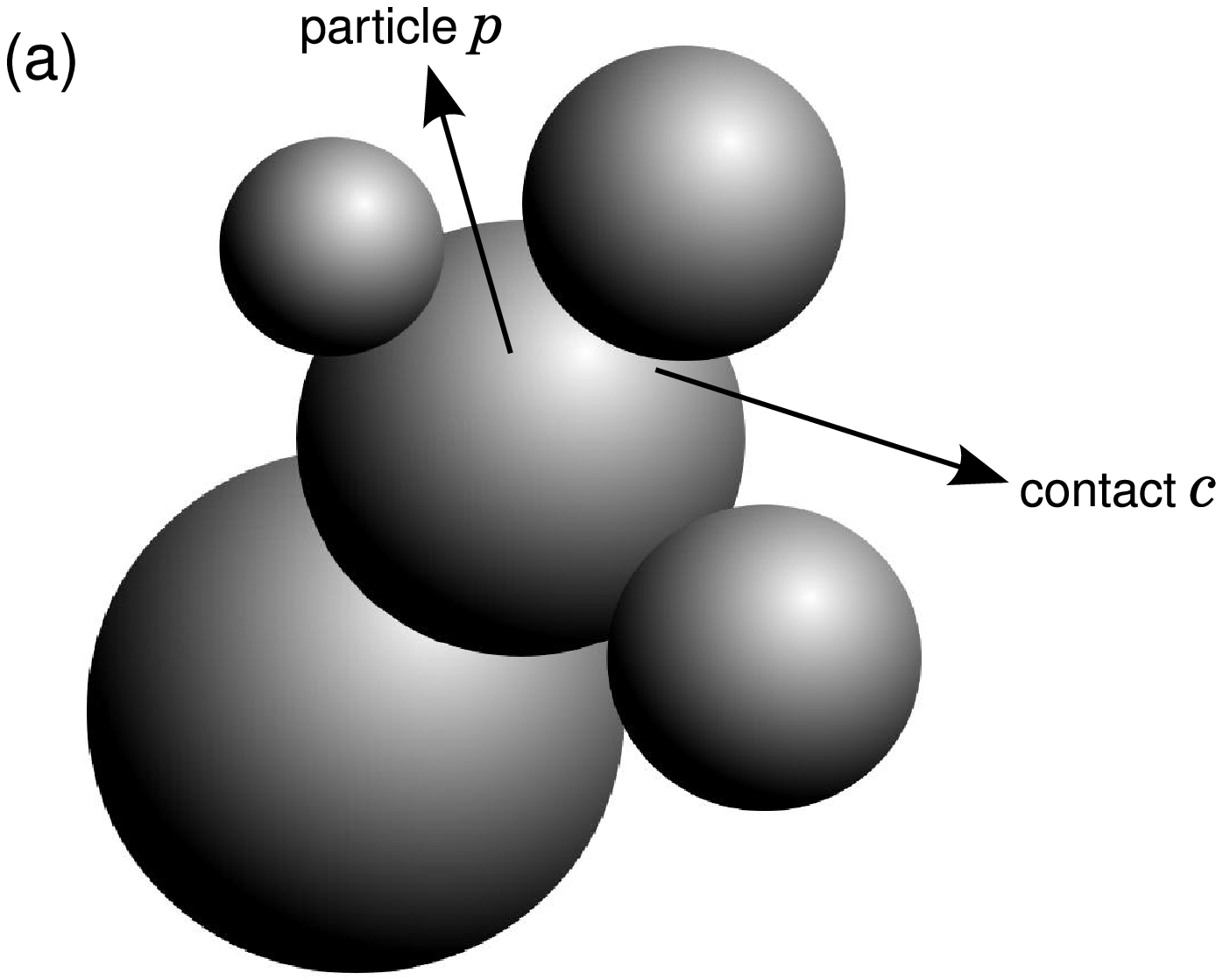}
\includegraphics[scale=0.30,angle=-90]{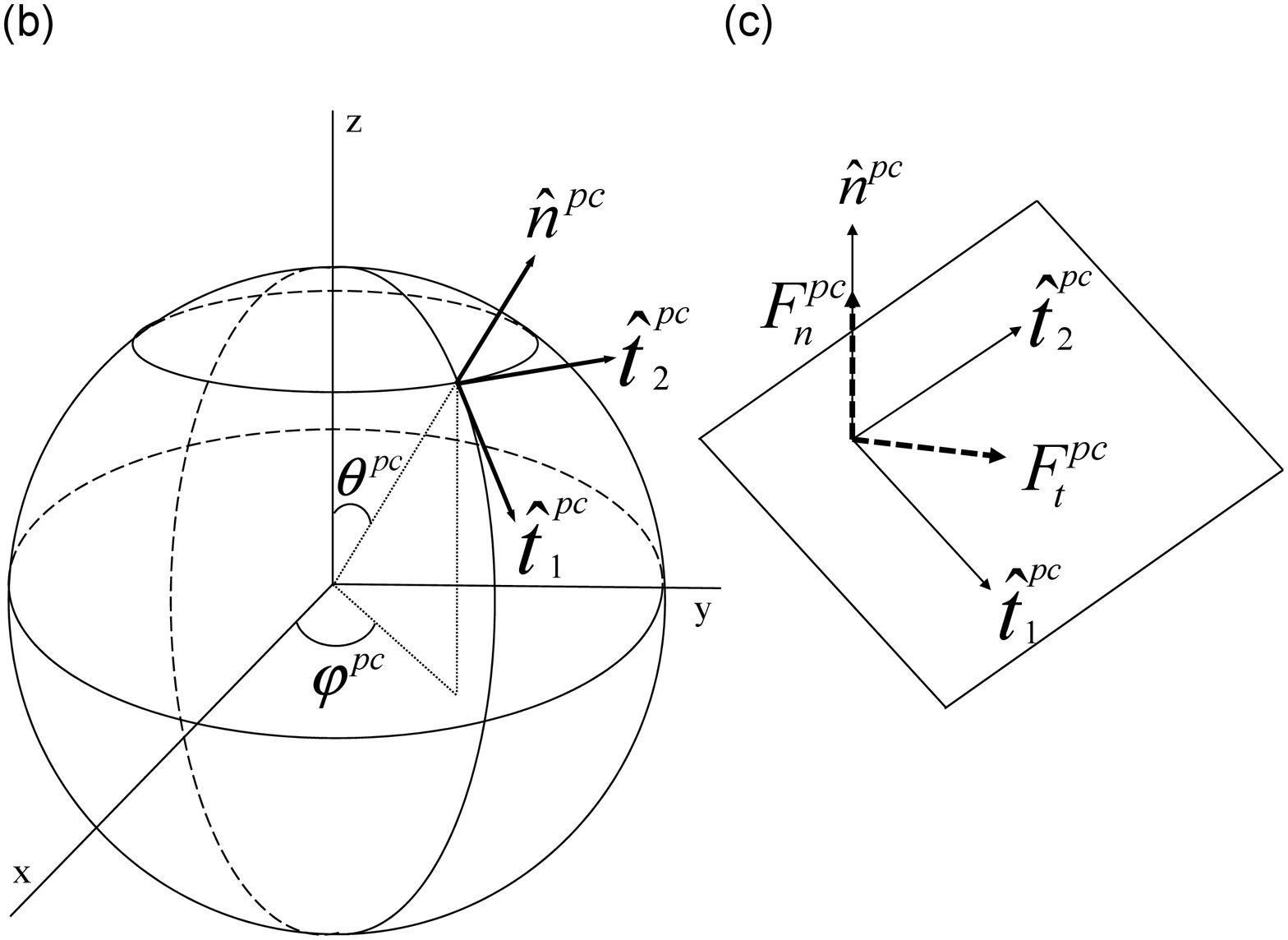}
\caption{(a) A typical contact $c$ between the reference particle
$p$ and its neighboring particle. (b) The contact unit vectors
$\hat n^{pc}$, $\hat t_1^{pc}$, and $\hat t_2^{pc}$. (c) The normal
($F^{pc}_n$) and tangential ($F^{pc}_t$) components of the
contact force $\vec F^{^{pc}}\!\!\!$.} 
\label{Fig-Schematic2}
\end{figure}

The micromechanical expressions for the components of the stress
tensor $\sigma^p_{\alpha\beta}$ of a single particle in a static 
granular assembly are \cite{Christoffersen81,Latzel00}
\begin{equation}
 \sigma^{^p}_{_{\alpha \beta}} = \frac{1}{V_p}\sum_{c=1}^{C_p}
 l^{^{pc}}_\alpha F^{^{pc}}_\beta,
  % \null_{\text{stress-single}}
  \label{stress-single}
\end{equation}
where $\vec F^{^{pc}}$ is the force exerted on particle $p$ by its
neighboring particle at contact $c$. 

One could assume in a crude approximation that the force at 
contact $c$ is equal to $\bar F_n^{p} \hat n^{pc} + \bar F_{t_1}^{p} 
{\hat t}_{_1}^{pc}+ \bar F_{t_2}^{p} {\hat t}_{_2}^{pc}$ in a 
three-dimensional system, where $\bar F_n^{p}$, $\bar F_{t_1}^{p}$, and 
$\bar F_{t_2}^{p}$ are the average normal and tangential contact 
forces around the particle $p$, and $\hat n^{pc}$, 
${\hat t}_{_1}^{pc}$, and ${\hat t}_{_2}^{pc}$ are the normal 
and tangential unit vectors at contact $c$, respectively. Then 
the \emph{force-averaged} stress tensor becomes
\begin{equation}
\widetilde\sigma^{^p}_{_{\alpha \beta}} {=} \frac{a_p}{V_p} \biggl( 
\bar F_n^{p} \sum_{c=1}^{C_p} n^{^{pc}}_\alpha n^{^{pc}}_\beta +
\bar F_{t_1}^{p} \sum_{c=1}^{C_p} n^{^{pc}}_\alpha t_{1\beta}^{^{pc}} +
\bar F_{t_2}^{p} \sum_{c=1}^{C_p} n^{^{pc}}_\alpha t_{2\beta}^{^{pc}}
\biggr).
  % \null_{\text{stress-single-meanfield}}
  \label{stress-single-meanfield}
\end{equation}
For a spherical grain, we project the contact unit vectors ($\hat
n^{^{pc}}$, $\hat t_{_1}^{^{pc}}$, $\hat t_{_2}^{^{pc}}$) onto an
arbitrary Cartesian coordinate system
[Figs.~\ref{Fig-Schematic2}(a) and \ref{Fig-Schematic2}(b)], and 
write the force-averaged stress tensor of a single particle as
\begin{eqnarray}
&&\hspace{-1.3cm}\widetilde\sigma^{^p} = \frac{a_p}{V_p} \sum_{c=1}^{C_p} 
\Biggl[ \bar F_n^{p} \left( \begin{array}{ccc}
\;W_{2002} & \;W_{2011} & \;W_{1101} \\
\;W_{2011} & \;W_{2020} & \;W_{1110} \\
\;W_{1101} & \;W_{1110} & \;W_{0200} \end{array} \right) 
\nonumber\\ && \hspace{0.45cm} + \bar F_{t_1}^{p} 
\left( \begin{array}{ccc}
\;W_{1102} & W_{1111} & \!\!-W_{2001} \\
\;W_{1111} & W_{1120} & \!\!-W_{2010} \\
\;W_{0201} & W_{0210} & \!\!-W_{1100} \end{array}
\right) \nonumber\\ && \hspace{0.45cm} + \bar F_{t_2}^{p}
 \left( \begin{array}{ccc}
\!\!\!-W_{1011} & W_{1002} & \;\;\;\; 0 \\
\!\!\!-W_{1020} & W_{1011} & \;\;\;\; 0 \\
\!\!\!-W_{0110} & W_{0101} & \;\;\;\; 0 \end{array}\;
\right) \Biggr],
  % \null_{\text{stress-single-meanfield-2}}
  \label{stress-single-meanfield-2}
\end{eqnarray}
where the $W_{mnkl}$ function is defined as
\begin{equation}
W_{mnkl} = \sin^m(\theta_c) \cos^n(\theta_c) \sin^k(\varphi_c)
\cos^l(\varphi_c),
  % \null_{\text{W-definition}}
  \label{W-definition}
\end{equation}
with $0 {\leqslant} \theta_c {<} \pi$ and $0 {\leqslant} \varphi_c 
{<} 2\pi$. Using Eq.~(\ref{stress-single-meanfield}), the trace of 
the stress tensor becomes
\begin{eqnarray}
\widetilde\sigma^{^p}_{_{\alpha \alpha}} {=} \frac{a_p}{V_p} 
\sum_{c=1}^{C_p} \sum_{\alpha=1}^{3} \biggl( \!\! \bar F_n^{p} 
n^{^{pc}}_\alpha n^{^{pc}}_\alpha {+} \bar F_{t_1}^{p} n^{^{pc}}_\alpha 
t_{1\alpha}^{^{pc}} {+} \bar F_{t_2}^{p} n^{^{pc}}_\alpha 
t_{2\alpha}^{^{pc}} \!\! \biggr) \nonumber \\
& \hspace{-7.25cm} {=} \displaystyle\frac{a_p}{V_p} \!\! 
\displaystyle\sum_{c=1}^{C_p} \biggl( \!\! \bar F_n^{p} 
|\hat n^{^{pc}}\!|^{^2}  {+} \bar F_{t_1}^{p} \hat n^{^{pc}} 
\!\!\!\!\cdot \hat t_{1}^{^{pc}} {+} \bar F_{t_2}^{p} \hat 
n^{^{pc}} \!\!\!\!\cdot \hat t_{2}^{^{pc}} \!\! \biggr ) {=} 
\frac{a_p}{V_p} \bar F_n^{p} C_p. \nonumber \\
  % \null_{\text{stress-singletrace}}
  \label{stress-single-trace}
\end{eqnarray}
Equation (\ref{stress-single-trace}) remains valid also in the 2D 
case (see Appendix). As expected for isotropic packings, the trace of 
the stress tensor and therefore the isotropic pressure $P$ ($\!=\! 
\sigma_{\alpha\alpha} / 3$) do not depend on the tangential forces.

\subsection{Many-particle case}
\label{Stress-Many} In the many-particle case, the average stress
tensor in a given volume $V$ is defined as \cite{Latzel00}
\begin{equation}
\langle \sigma_{_{\alpha \beta}} \rangle_{_V} =
\frac{1}{V}\sum^N_{p=1} V_p \sigma^{^p}_{\alpha \beta} = \frac{1}{V}
\sum^N_{p=1} \sum^{C_p}_{c=1} l^{^{pc}}_\alpha F^{^{pc}}_\beta,
  % \null_{\text{stress-many-1}}
  \label{stress-many-1}
\end{equation}
where the sum runs over all particles lying inside $V$. Using
Eq.~(\ref{stress-single-trace}) to calculate the trace of the
average stress tensor, we get
\begin{equation}
\langle \widetilde\sigma_{_{\alpha \alpha}} \rangle_{_V} =
\frac{1}{V}\sum^N_{p=1} V_p \widetilde\sigma^{^p}_{\alpha 
\alpha} = \frac{1}{V} \sum^N_{p=1} a_p \bar F^{^p}_n C_p.
  % \null_{\text{stress-many-2}}
  \label{stress-many-2}
\end{equation}

\subsection{Polydispersity}
\label{Stress-Polydispersity}

Now we assume a polydisperse distribution of particle radii with
probability $f(a) da$ to find the radius between $a$ and $a+da$,
and with $\int_{_0}^{^\infty} \!\!f(a)da=1$. Assuming that the 
average contact force exerted on a particle depends only on its 
radius $a$, the continuous limit of Eq.~(\ref{stress-many-2}) in 
a mean-field approximation is given by
\begin{equation}
\langle \widetilde\sigma_{\alpha \alpha} \rangle_{_V} = \frac{N}{V}
\int_0^\infty a \bar F_n(a) C(a) f(a) da.
  % \null_{\text{stress-many-continuous}}
  \label{stress-many-continuous}
\end{equation}
In Eq.~(\ref{stress-many-continuous}), it is supposed that all
particles of size $a$ have a certain mean coordination number
$C(a)$ and a certain mean normal force $\bar F_n(a)$. Indeed, 
particles of the same size may have different coordination
number and normal contact forces, however, the main goal here is
to propose a method to calculate macroscopic quantities without
taking into account all the microscopic details of the system. We
use the mean-field approach introduced in
Sec.~\ref{Fabric-Polydispersity} to evaluate $C(a)$. By
substitution of Eq.~(\ref{coordination-number}) in
Eq.~(\ref{stress-many-continuous}) we get
\begin{eqnarray}
\langle \widetilde\sigma_{_{\alpha \alpha}} \rangle_{_V} = \frac{N}{V} z
\frac{\int_0^\infty a \bar F_n(a) \frac{f(a)}{\Omega(a)}
da}{q_{_0}} \hspace{2.5cm} \nonumber \\
& \hspace{-9.0cm} = \phi z \displaystyle\frac{\displaystyle\int_0^\infty a \bar F_n(a)
\frac{f(a)}{\Omega(a)} da}{q_{_0}\displaystyle\int_0^\infty V(a) f(a) da}.
  % \null_{\text{stress-poly-1}}
  \label{stress-poly-1}
\end{eqnarray}
According to the mean-field approach used in Sec.~\ref{Fabric-Polydispersity},
$C(a)$ increases with increasing radius $a$. Now, let us
assume that the average normal force $\bar F_n(a)$ also increases
with $a$, so that the ratio $\bar F_n(a)/C(a)$ remains roughly
constant \cite{Madadi05}. We calculate this ratio for the 
average-sized particles in the following:
\begin{equation}
\frac{\bar F_n(a)}{C(a)} = \frac{\bar F_n(a)}{\frac{z}
{q_{_0}\Omega(a)}} \simeq \frac{\bar F_n(\langle a \rangle)}
{\frac{z}{q_{_0}\Omega(\langle a \rangle)}} =
\frac{q_{_0}\bar F_n(\langle a \rangle)\Omega(\langle a
\rangle)}{z},
  % \null_{\text{stress-poly-3}}
 \label{stress-poly-3}
\end{equation}
therefore
\begin{equation}
\bar F_n(a) = \frac{\Omega(\langle a \rangle) \bar F_n(\langle a 
\rangle)}{\Omega(a)}.
  % \null_{\text{stress-poly-4}}
 \label{stress-poly-4}
\end{equation}
By substitution of Eq.~(\ref{stress-poly-4}) in Eq.~(\ref{stress-poly-1}), 
we obtain
\begin{eqnarray}
\langle \widetilde\sigma_{_{\alpha \alpha}} \rangle_{_V} = 
\displaystyle\frac{3 \phi z \; \bar F_n(\langle \displaystyle{a} \rangle) 
\; g_{_2}}{4 \pi \big\langle \displaystyle{a}^2 \big\rangle}
% \null_{\text{stress-poly-5}}
\label{stress-poly-5}
\end{eqnarray}
with
\begin{eqnarray}
g_{_2} = \displaystyle\frac{(2{-}\sqrt{3})\pi \,\langle 
\displaystyle{a}^2 \rangle \!\displaystyle\int_0^\infty \!\!\!\! 
\displaystyle{a} \frac{f(\displaystyle{a})} {\Omega^2(\displaystyle{a})} 
d\displaystyle{a} }{q_{_0} \langle \displaystyle{a}^3 \rangle}
% \null_{\text{stress-poly-6}}
\label{stress-poly-6}
\end{eqnarray}

\subsection{Narrow size distributions}
\label{Stress-NarrowDistribution}

In the limit of narrow size distributions, we approximate
$1/\Omega^2(a)$ by Taylor expansion around $\epsilon = 0$
(similar to Sec.~\ref{Fabric-NarrowDistribution}):
\begin{equation}
\frac{1}{\Omega^2(a)} \simeq A_{_2} + B_{_2}\epsilon +
C_{_2}\epsilon^2,
  % \null_{\text{space-angle-sqr-expansion}}
  \label{space-angle-sqr-expansion-3D}
\end{equation}
with $A_{_2} \!\!=\! A_{_1}^{^2} \!\!=\! \frac{1}{(2-\sqrt 3)^2 \pi^2 }$, $B_{_2}
\!\!=\! \frac{1}{\sqrt 3 (2-\sqrt 3)^3 \pi^2}$, and $C_{_2}
\!\!=\! \frac{1}{4(2-\sqrt 3)^4\pi^2} - \frac{5 \sqrt
3}{18(2-\sqrt 3)^3\pi^2}$. By substitution of
Eq.~(\ref{space-angle-sqr-expansion-3D}) in Eq.~(\ref{stress-poly-6}) 
we obtain the correction factor $g_{_2}$ for arbitrary narrow distributions,
\begin{equation}
g_{_2} \!\! \simeq \!\!\frac{(A_{_2}{-}B_{_2}{+}C_{_2})\frac{\big\langle 
\displaystyle{a} \big\rangle \big\langle \displaystyle{a}^2 
\big\rangle}{\big\langle \displaystyle{a}^3 \big\rangle} {+} 
(B_{_2}{-}2C_{_2}) \frac{\big\langle \displaystyle{a}^2 
\big\rangle^2}{\big\langle \displaystyle{a} \big\rangle 
\big\langle \displaystyle{a}^3 \big\rangle} {+} 
C_{_2} \frac{\big\langle \displaystyle{a}^2 
\big\rangle}{\big\langle \displaystyle{a} \big\rangle^2}}
{(A_{_2}{-}A_{_1}C_{_1}){+}A_{_1}C_{_1} \frac{\big\langle 
\displaystyle{a}^2 \big\rangle}{\big\langle \displaystyle{a} 
\big\rangle^2}}.
% \null_{\text{g1-narrow}}
\label{g1-narrow}
\end{equation}

\section{Stiffness tensor}
\label{Stiffness-Tensor}

The linear response of a material to ``weak" external 
perturbations is described by a fourth rank tensor, which 
is called the \emph{elastic} or \emph{stiffness} tensor 
\cite{Kruyt96,Luding04}. This tensor has $81$ and $16$ 
elements in three- and two-dimensional systems, respectively, 
but they are not all independent. Symmetry considerations 
reduce the number of independent elements. For example, the 
elastic behavior of isotropic materials can be described by 
only two independent parameters, usually represented by Lam\'e 
coefficients $\lambda$ and $\mu$. In this section, the 
stiffness tensor of a homogeneous and isotropic assembly 
of polydisperse particles is investigated (for the case of 
an anisotropic monodisperse system, see, e.g.\ \cite{Shaebani11,
Mouraille06}).

The stiffness tensor for a spherical particle, where affine 
deformation is assumed, is defined as 
\cite{Bathurst88,Luding04} 
\begin{equation}
\mathcal{C}^{p}_{\alpha,\beta,\gamma,\eta} {=} \frac{2 a_p^2} 
{V_p} \sum^{C_p}_{c=1} (k_n n^{pc}_{\alpha} n^{pc}_{\beta} 
n^{pc}_{\gamma} n^{pc}_{\eta} + k_t n^{pc}_{\alpha} 
t^{pc}_{\beta} n^{pc}_{\gamma} t^{pc}_{\eta}),
  % \null_{\text{stiffness-1}}
  \label{stiffness-1}
\end{equation}
where $\hat t^{pc}$ is the unit vector parallel to the 
tangential component of the contact force $\vec F^{pc}$ [see 
Fig.~\ref{Fig-Schematic2}(c)]. The volume weighted average of 
$\mathcal{C}$ is then given by
\begin{eqnarray}
\langle \mathcal{C}_{\alpha,\beta,\gamma,\eta} \rangle_{_V} {=} 
\frac{1}{V} \sum^N_{p=1} V_p \mathcal{C}_{\alpha,\beta,\gamma,\eta}^p 
{=} \nonumber \\
& \hspace{-4.9cm} \displaystyle\frac{1}{V} \displaystyle\sum^N_{p=1} 
2 a_p^2 \displaystyle\sum^{C_p}_{c=1} (k_n n^{pc}_{\alpha} 
n^{pc}_{\beta} n^{pc}_{\gamma} n^{pc}_{\eta} {+} k_t n^{pc}_{\alpha} 
t^{pc}_{\beta} n^{pc}_{\gamma} t^{pc}_{\eta}).
  % \null_{\text{stiffness-2}}
  \label{stiffness-2}
\end{eqnarray}
Note that the stiffness tensor is basically determined by the 
packing geometry. For ease of calculation, we consider only 
frictionless packings, i.e.\ $k_t$ is set to zero hereafter. 
Using the microscopic information of the contact orientations, 
one can accurately calculate the elements of $\mathcal{C}$ via 
Eq.~(\ref{stiffness-2}). Next, the Lam\'e constants $\mu$ and 
$\lambda$ can be deduced from the stiffness tensor, e.g.\ as 
$\lambda {=} \langle \mathcal{C}_{_{\!1122}} \rangle_{_V}$ and 
$\lambda {+} 2 \mu {=} \langle \mathcal{C}_{_{\!1111}} 
\rangle_{_V}$ or, more generally, as $\lambda {=} \langle 
\overline {\mathcal{C}_{_{\!iijj}}} \rangle_{_V}$ and $\lambda 
{+} 2 \mu {=} \langle \overline {\mathcal{C}_{_{\!iiii}}}\rangle_{_V}$ 
where 
\begin{equation} 
\langle \overline {\mathcal{C}_{_{\!iijj}}} \rangle_{_V} {=} \frac{1}{D(D{-}1)} 
\sum_{i \neq j}^D \langle \mathcal{C}_{_{\!iijj}} \rangle_{_V} \; , \; \langle 
\overline {\mathcal{C}_{_{\!iiii}}} \rangle_{_V} {=} \frac{1}{D} 
\sum_i^D \langle \mathcal{C}_{_{\!iiii}} \rangle_{_V},
\label{average-C-elements}
\end{equation}
and $D$ is the dimension of the system. The macroscopic physical 
quantities of interest are the bulk modulus $K$ and the shear 
modulus $G$, which can be deduced from the Lam\'e coefficients in 
isotropic materials as
\begin{eqnarray}
\hspace{-2.0cm} G/k_n {=} \mu/k_n {=} \frac{\langle \overline {\mathcal{C}_{_{\!iiii}}} 
\rangle_{_V}{-}\langle \overline {\mathcal{C}_{_{\!iijj}}} \rangle_{_V}} 
{2 \, k_n},
% \null_{\text{Shear-modulus}}
\label{Shear-modulus}
\end{eqnarray}
and
\begin{eqnarray}
K/k_n {=} (\lambda {+} \frac{2}{D} \mu)/k_n {=} \frac{\langle 
\overline {\mathcal{C}_{_{\!iiii}}} \rangle_{_V} {+} (D {-} 1) 
\langle \overline {\mathcal{C}_{_{\!iijj}}} \rangle_{_V}}{D \, k_n}.
% \null_{\text{Bulk-modulus}}
\label{Bulk-modulus}
\end{eqnarray}

Now, assuming a polydisperse probability distribution of 
particle radii $f(a)$, Eq.~(\ref{stiffness-2}) for 
$k_t{=}0$ can be written as
\begin{equation}
\langle \mathcal{C}_{\alpha,\beta,\gamma,\eta} \rangle_{_V} 
{=} \frac{N k_n}{V} \int_0^\infty 2a^2 \bigg ( 
\displaystyle\sum^{C(a)}_{c=1} n^{c}_{\alpha} n^{c}_{\beta} 
n^{c}_{\gamma} n^{c}_{\eta}\bigg ) f(a) da.
  % \null_{\text{stiffness-3}}
  \label{stiffness-3}
\end{equation}
Since the packings are supposed to be isotropic and homogeneous,
we assume that grains are scattered homogeneously around the
reference particle. Therefore, the summation over neighbors can
be approximated by the following integration in three dimensions 
(for the 2D case, see Appendix):
\begin{equation}
\sum^{C(a)}_{c=1} Q(\theta^{c},\varphi^{c}) = \frac{C(a)}{4\pi}
\int_0^\pi \!\!\!\! d\theta \; \sin(\theta) \int_0^{2\pi} \!\!\!\! 
d\varphi \; Q(\theta,\varphi).
  % \null_{\text{Isotropic-average}}
  \label{Isotropic-average}
\end{equation}
We present the reduced form of the fourth rank tensor by mapping 
$\alpha\beta(\gamma\eta) \rightarrow i(j)$, i.e.\ $11 
\rightarrow 1$, $22 \rightarrow 2$, $33 \rightarrow 3$, $12 
\rightarrow 4$, $13 \rightarrow 5$, and $23 \rightarrow 6$. 
Using Eqs.~(\ref{coordination-number}), (\ref{stiffness-3}), and 
(\ref{Isotropic-average}), one obtains
\begin{eqnarray}
\langle \mathcal{C} \rangle_{_V} {=} \frac{N k_n z}{2 \pi V q_{_0}} 
\!\! \int_0^\infty \!\!\!\!\!\! da \; \frac{a^2}{\Omega(a)} f(a) 
{\times} \nonumber \\
& \hspace{-5.0cm} \displaystyle\int_0^\pi \!\!\!\! d\theta \; 
\sin(\theta) \displaystyle\int_0^{2\pi} \!\!\!\!\!\! d\varphi 
\!\!\left( \begin{array}{ccccccccc}
\!\!W_{_{\!4004}} & \!\!\!\!W_{_{\!4022}} & \!\!\!\!W_{_{\!2202}} &
\!\!\!\!W_{_{\!4013}} & \!\!\!\!W_{_{\!3103}} & \!\!\!\!W_{_{\!3112}} \\
& \!\!\!\!W_{_{\!4040}} & \!\!\!\!W_{_{\!2220}} & \!\!\!\!W_{_{\!4031}}
& \!\!\!\!W_{_{\!3121}} & \!\!\!\!W_{_{\!3130}} \\
& & \!\!\!\!W_{_{\!0400}} & \!\!\!\!W_{_{\!2211}} &
\!\!\!\!W_{_{\!1301}} & \!\!\!\!W_{_{\!1310}} \\
& & & \!\!\!\!W_{_{\!4022}} & \!\!\!\!W_{_{\!3112}} &
\!\!\!\!W_{_{\!3121}} \\
& & & & \!\!\!\!W_{_{\!2202}} & \!\!\!\!W_{_{\!2211}} \\
& & & & & \!\!\!\!W_{_{\!2220}} \end{array} \!\! \right), \nonumber \\
  % \null_{\text{stiffness-4}}
  \label{stiffness-4}
\end{eqnarray}
where $W_{_{\!ijkl}}$ elements were defined in Eq.~(\ref{W-definition}). 
After integration on $\theta$ and $\varphi$, the volume weighted
average of the stiffness tensor for an isotropic polydisperse 
packing becomes
\begin{equation}
\langle \mathcal{C} \rangle_{_V} =
\frac{\phi z k_n g_{_3}}{10\pi \langle \displaystyle{a} \rangle} \left(
\begin{array}{ccccccccc}
3 & 1 & 1 & 0 & 0 & 0 \\
& 3 & 1 & 0 & 0 & 0 \\
& & 3 & 0 & 0 & 0 \\
& & & 1 & 0 & 0 \\
& & & & 1 & 0 \\
& & & & & 1 \end{array} \right).
  % \null_{\text{stiffness-5}}
  \label{stiffness-5}
\end{equation}
The correction factor $g_{_3}$ is defined as
\begin{equation}
g_{_3} = \frac{\langle \displaystyle{a} \rangle \langle 
\displaystyle{a}^2 \rangle_{_g}}{\langle \displaystyle{a}^3 \rangle},
% \null_{\text{stiffness-7}}
\label{stiffness-7}
\end{equation}
and for narrow size distributions, one obtains
\begin{equation}
g_{_3} \!\simeq\! \frac{\big(A_{_1} {-} B_{_1} {+} C_{_1}\big) 
\frac{\big\langle \displaystyle{a} \big\rangle \big\langle \displaystyle{a}^2 
\big\rangle}{\big\langle \displaystyle{a}^3 \big\rangle} {+} \big(B_{_1} {-} 
2C_{_1}\big) {+} C_{_1}\frac{\big\langle \displaystyle{a}^4 \big\rangle} 
{\big\langle \displaystyle{a} \big\rangle \big\langle \displaystyle{a}^3 
\big\rangle}}{\big(A_{_1} {-} C_{_1}\big) {+} C_{_1} \frac{\big\langle 
\displaystyle{a}^2 \big\rangle}{\big\langle \displaystyle{a} \big\rangle^2}},
% \null_{\text{stiffness-8}}
\label{stiffness-8}
\end{equation}
with the same coefficients as defined after 
Eqs.~(\ref{space-angle-expansion-3D}) and 
(\ref{space-angle-sqr-expansion-3D}). To summarize this section, the 
Lam\'e constants for frictionless isotropic packings are
\begin{equation}
\mu = \lambda = (k_n \phi z g_{_3}) \, / \,(10\pi \langle a \rangle ),
  % \null_{\text{stiffness-9}}
  \label{stiffness-9}
\end{equation}
and the shear and bulk moduli are
\begin{equation}
G/k_n = (\phi z g_{_3}) \, / \,(10\pi \langle a \rangle),
  % \null_{\text{stiffness-10}}
  \label{stiffness-10}
\end{equation}
and
\begin{equation}
K/k_n = (\phi z g_{_3}) \, / \, (6\pi \langle a \rangle).
  % \null_{\text{stiffness-11}}
  \label{stiffness-11}
\end{equation}
Notably, $K/G \,{=}\, 5/3$ in three-dimensional frictionless isotropic 
packings, independent of their size distribution and average packing 
properties.

\section{Simulation results}
\label{Simulation-Results}

\begin{figure}[b]
\centering
\includegraphics[scale=0.32,angle=0]{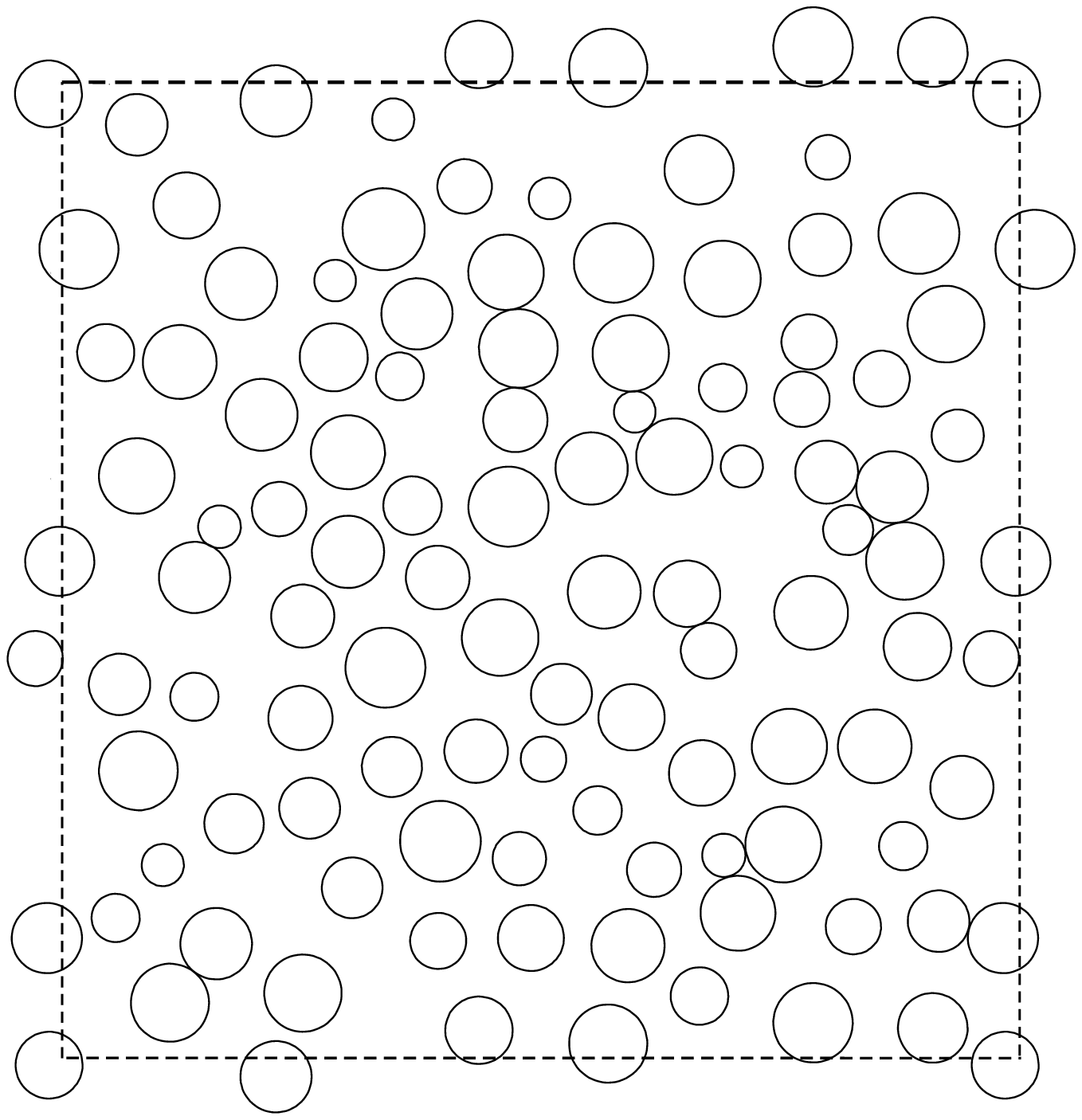}
\includegraphics[scale=0.32,angle=0]{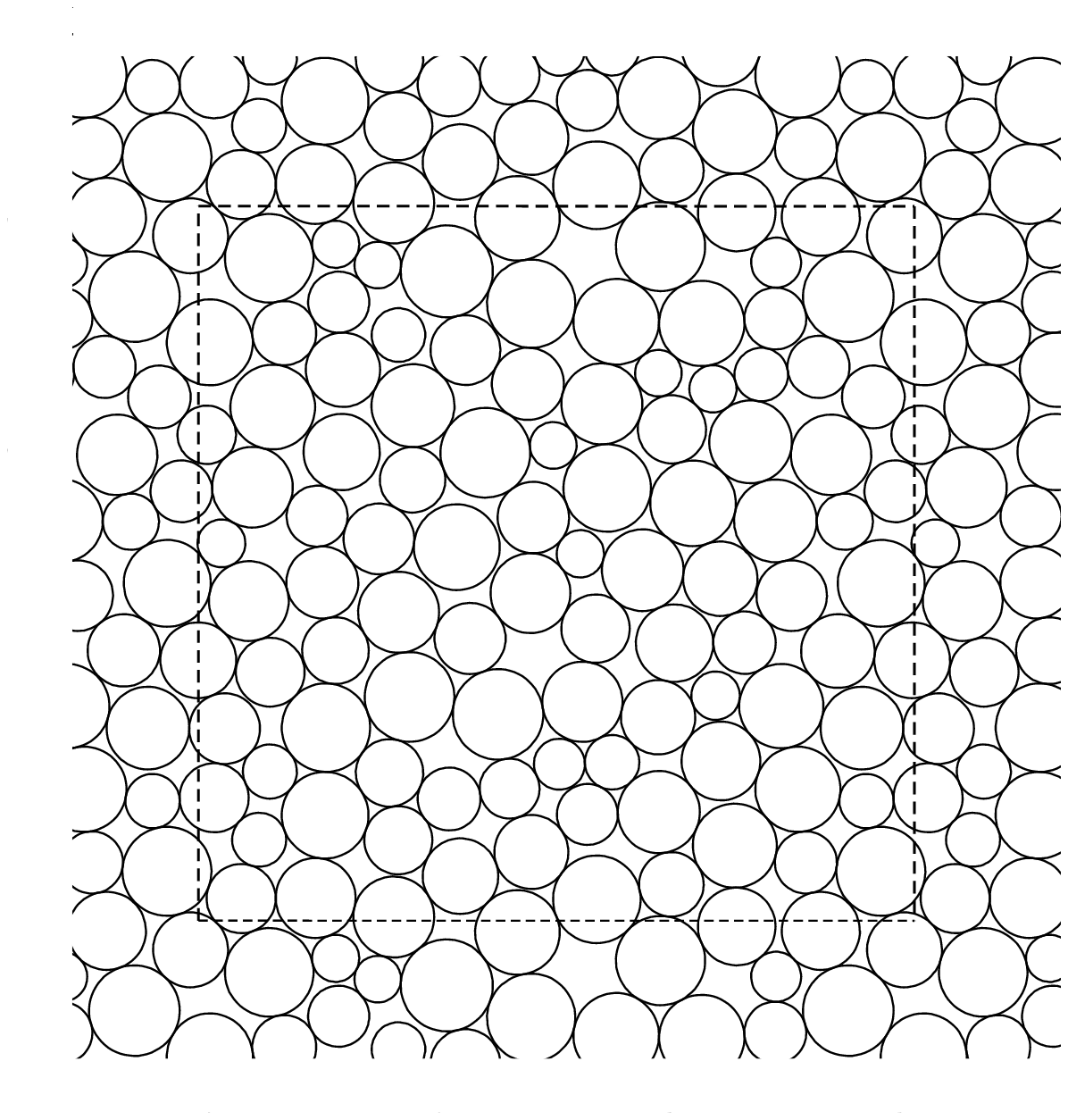}
\caption{Schematic of a 2D granular system subjected to a
constant external pressure: the initial dilute gas (left), 
and the final homogeneous packing (right). Periodic boundaries 
are marked with dashed lines.} \label{Fig-Schematic3}
\end{figure}

To verify the theoretical predictions of the previous sections,
we carry out numerical simulations with the help of the contact
dynamics (CD) algorithm \cite{Moreau94,Jean99,Brendel04}. We
first construct 2D and 3D static homogeneous packings in zero
gravity by compressing the initial dilute configuration of 
particles [Fig.~\ref{Fig-Schematic3} (left)]. Periodic boundary 
conditions are imposed in all directions to avoid the side effects 
of lateral walls. The compaction is achieved by imposing a 
constant external pressure $P_\text{ext}$ and letting the size 
of the system evolve in time \cite{Andersen80}. As the volume of 
the system decreases, after a while particles touch each other 
and build an inner pressure $P_\text{inn}$, which resists and 
eventually compensates $P_\text{ext}$, so that finally $P_\text{inn}$ 
equals $P_\text{ext}$. Particles prevent further compaction, 
and a static homogeneous configuration is reached 
[Fig.~\ref{Fig-Schematic3} (right)]. The full description of the 
packing generation method can be found in \cite{Shaebani09}. In 
order to illustrate the validity range of our assumptions, we generate 
three types of polydisperse packings with uniform particle-size 
distributions but with different widths (see Table~\ref{Table1}). 
We denote the samples SMP1, SMP2, and SMP3, respectively, by full 
circles, open squares, and full triangles throughout this section.
To investigate the effect of friction, we construct a new 
packing for each value of the particle-particle friction 
coefficient $\mu_{_f}$. In particular, the results corresponding to 
$\mu_{_f}{=}0$, $0.1$, and $1.0$ are hereafter denoted by green, 
blue and red colors. The number of grains contained by packings 
are $3000$ and $10000$ in 2D and 3D cases, respectively.

\begin{table}[t]
\caption{Properties of three different types of polydisperse 
packings generated with uniform size distributions. $w$ denotes 
the width of each distribution ($w = a_{_{max}}\!{-}a_{_{min}}$).}
\begin{center}
\begin{tabular}{ccccccccc}
\hline
type  & symbol && $a_{_{min}}$ && $a_{_{max}}$ & $a_{_{max}}/
a_{_{min}}$ & $w/2\langle a \rangle$ & $\langle a^2 \rangle / 
\langle a \rangle^2 $ \\
SMP1  & $\bullet$ && 0.67 && 1.34 & 2 & 0.34 & 1.04 \\
SMP2  & $\square$ && 0.40 && 1.60 & 4 & 0.60 & 1.12 \\
SMP3  & $\blacktriangle$ && 0.22 && 1.76 & 8 & 0.77 & 1.19 \\
\hline 
\label{Table1}
\end{tabular}
\end{center}
\end{table}

\begin{figure}[b]
\centering
\includegraphics[scale=0.8,angle=0]{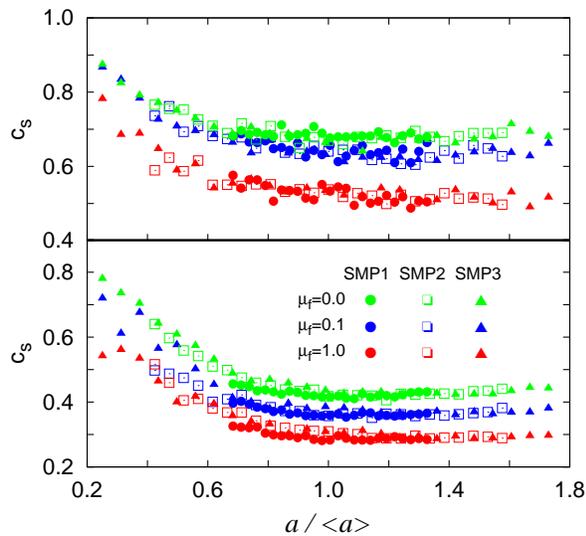}
\caption{(Color online) Linear compacity $c_s$ as a function of 
particle radius $a$ for two-dimensional (top) and three-dimensional 
(bottom) packings constructed with different size distribution widths 
and different friction coefficients $\mu_{_f}$.}
\label{Fig-Compacity}
\end{figure}

\begin{figure}[t]
\centering
\includegraphics[scale=0.8,angle=0]{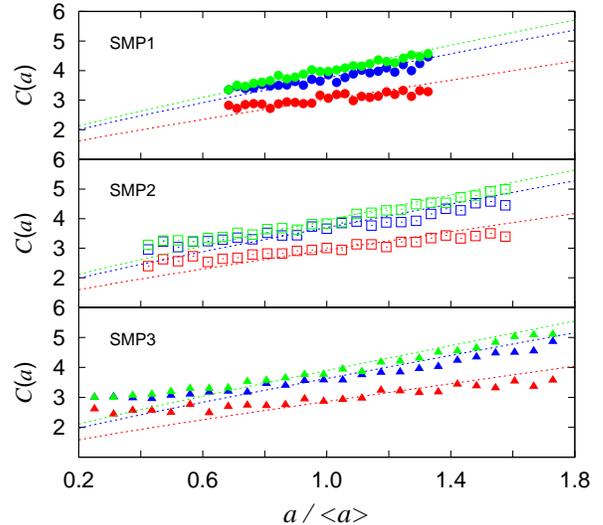}
\caption{(Color online) Contact number $C(a)$ as a function of 
particle radius $a$ for two-dimensional packings. The lines 
correspond to the mean-field approximation of $C(a)$ according 
to Eq.~(\ref{coordination-number}).}
\label{Fig-Coordination-2D}
\end{figure}

\begin{figure}[b]
\centering
\includegraphics[scale=0.8,angle=0]{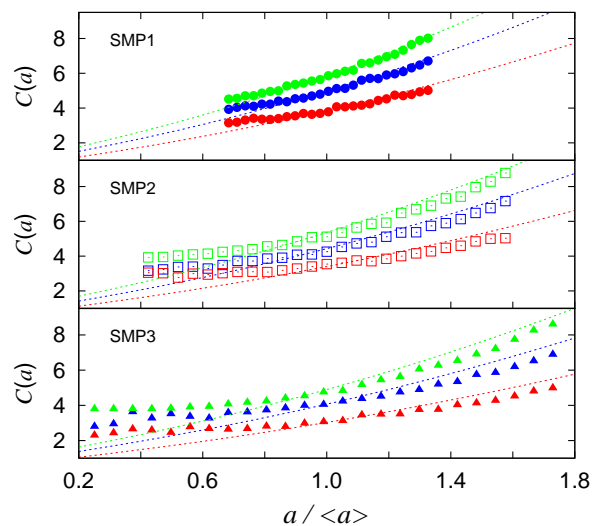}
\caption{(Color online) The same plots as in 
Fig.~\ref{Fig-Coordination-2D}, but for three-dimensional packings.}
\label{Fig-Coordination-3D}
\end{figure}

\begin{figure*}[t]
\centering
\includegraphics[scale=0.70,angle=0]{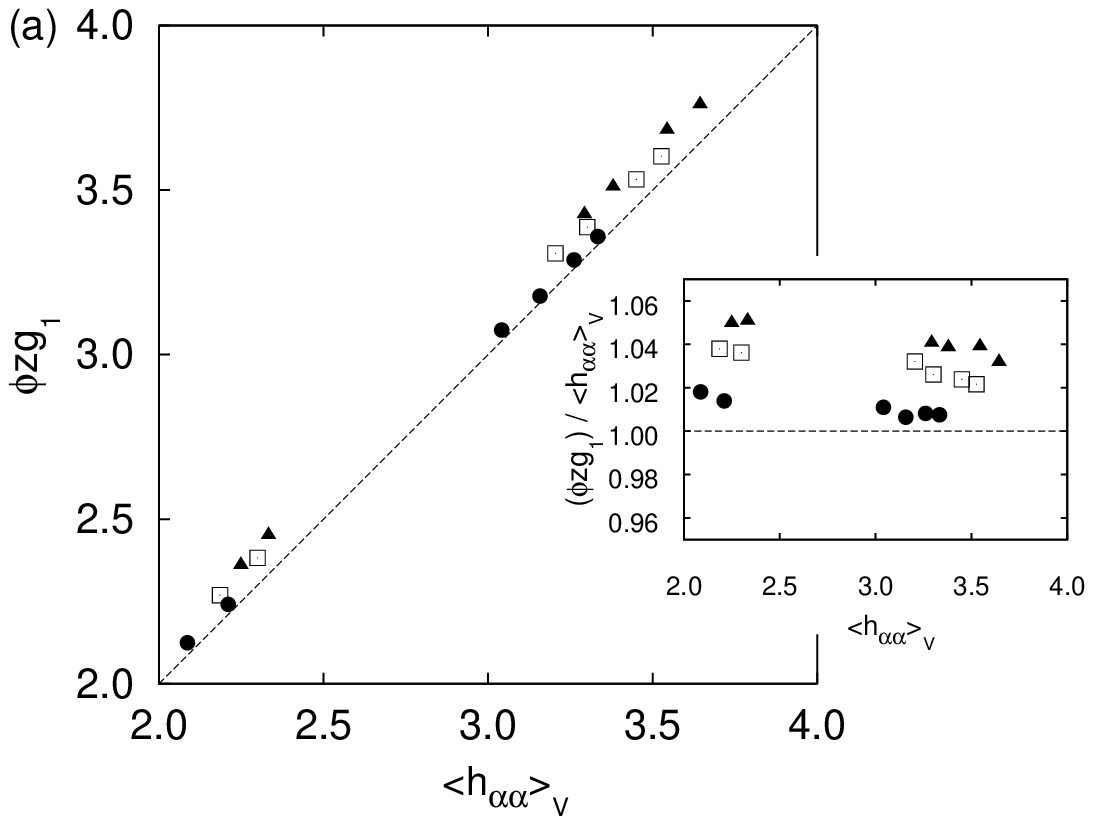}
\includegraphics[scale=0.70,angle=0]{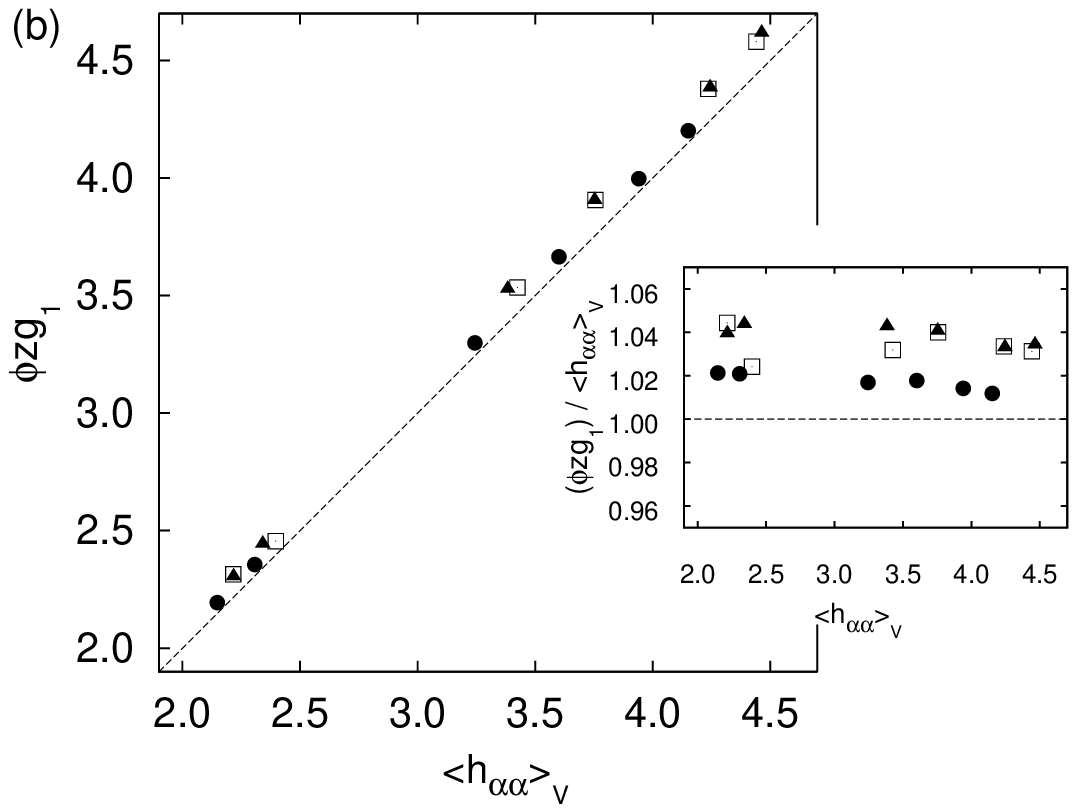}
\caption{The estimated value of the trace of the average fabric
tensor $\phi z g_{_1}$ vs the exact value $\langle
{\mathit h}_{_{\alpha \alpha}} \rangle_{_V}$ obtained from
the simulations, for several (a) two- and (b) three-dimensional
samples. The dashed lines indicate the identity. The insets show
more precisely that the deviations increase with $w$, but remain 
less than $5\%$ in all cases. The symbols are chosen the same as 
in Table~\ref{Table1}.} 
\label{Fig-fabric}
\end{figure*}

For comparison with the theory, we first test the validity of
assumptions made in Sec.~\ref{Fabric-Polydispersity}. The linear
compacity $c_s$ is displayed in Fig.~\ref{Fig-Compacity} 
for the static configurations of particles obtained from the 
isotropic compression simulations. For each particle $p$, the surface angle 
$\Omega^{^p}_c$ covered by its neighboring particle at contact $c$ 
is calculated, and the linear compacity of particle $p$ is obtained 
as $c_s^{^p} {=} \sum_{c=1}^{C_p} \Omega^{^p}_c/2\pi$ or $c_s^{^p} {=} 
\sum_{c=1}^{C_p} \Omega^{^p}_c/4\pi$ for two- or three-dimensional 
packings, respectively. Next, we divide the 
range of possible values of the particle radius $a$ into $25$ bins. 
Each data point in Fig.~\ref{Fig-Compacity} corresponds to the 
mean value of $c_s$, averaged over all particles in the same bin. 
The contribution of the rattler particles, which transmit no force, 
is excluded. For moderate widths of size distributions (SMP1), 
$c_s$ is approximately constant in $a$ for a given packing (we 
note that the fluctuations of $c_s$ around its mean value in a 
given packing originate from the finite size of the samples). 
However, $c_s$ is remarkably above the average value for small 
particle sizes in wider distributions (SMP2 and SMP3). This is 
a common property of our highly polydisperse packings (with 
uniform size distribution) that the fraction of shielded surface 
is larger than the average for small particles if rattlers are 
excluded (see \cite{Ogarko12} for uniform volume distributions). 
A similar behavior has been observed in discrete element method 
simulations of soft particles \cite{Goncu10}. There, it is also 
shown that if rattlers are included in the statistics, the small 
particles on average are less covered than the larger ones. However, 
the deviation of small particles from the average $c_s$ decreases 
as the volume fraction of the packing increases by incremental 
compression.

Another point is that $c_s$ depends strongly on the dimension of 
the system and the friction coefficient. Increasing the friction 
$\mu_{_f}$ stabilizes the system in a less dense state and decreases 
the connectivity of the contact network \cite{Kadau03,Shaebani09b}. 
Therefore, we expect lower values of $c_s$ and $C(a)$ when increasing 
$\mu_{_f}$, as confirmed by the data.

In Figs.~\ref{Fig-Coordination-2D} and \ref{Fig-Coordination-3D}, 
the coordination number $C(a)$ is shown as a function of $a$ for 
the same set of systems as in Fig.~\ref{Fig-Compacity}. For 
comparison, we also plot $C(a)$ from Eq.~(\ref{coordination-number}). 
Here, the average coordination number $z$ of the packing is taken 
from the simulation results, $\Omega(a)$ is provided by 
Eq.~(\ref{space-angle}) or (\ref{App-1}), and the size distribution 
of each packing after the compaction process is used to calculate 
$q_{_0}$. The mean-field approach of Sec.~\ref{Fabric-Polydispersity} 
qualitatively fits well to the data, however, the slopes of the curves 
are slightly greater than the corresponding slopes of the best-fit 
curves over the data points (not shown). Consequently, one expects 
that the mean-field approach to calculate the trace of the fabric 
tensor $\langle {\mathit h}_{_{\alpha \alpha}} \rangle_{_V}$ leads 
to somewhat overestimated values. For each packing, we calculate 
the exact value of $\langle {\mathit h}_{_{\alpha \alpha}}
\rangle_{_V}$ via Eq.~(\ref{fabric-many-trace}) and compare it
with the mean-field approximation [Eq.~(\ref{fabric-meanfield})].
Figure \ref{Fig-fabric} reveals that Eq.~(\ref{fabric-meanfield})
slightly overestimates $\langle {\mathit h}_{_{\alpha \alpha}}
\rangle_{_V}$ in both two- and three-dimensional systems. The 
deviation increases with the width of the size distribution, but 
remains less than $5\%$ in all cases. For comparison, note that 
$g_{_1}$ can reach up to $1.19$ and $1.45$ in 2D and 3D uniform 
samples, respectively (see Table~\ref{Table2}); therefore, 
ignoring the correction factor would cause up to $19\%$ and 
$45\%$ error, respectively.

\begin{table}[b]
\caption{ Correction factors in two and three dimensions for 
uniform size distributions SMP1, SMP2, and SMP3 introduced in 
Table \ref{Table1}.}
\begin{tabular}{cccccccccccccccccccccccccccc}
\hline
sample &&&&&&&&& $g_{_1}$ &&&&&&&&& $g_{_2}$ &&&&&&&&& $g_{_3}$\\
SMP1-2D  &&&&&&&&& 1.04 &&&&&&&&& 1.01 &&&&&&&&& 1.04 \\
SMP2-2D  &&&&&&&&& 1.12 &&&&&&&&& 1.04 &&&&&&&&& 1.12 \\
SMP3-2D  &&&&&&&&& 1.19 &&&&&&&&& 1.07 &&&&&&&&& 1.19 \\
SMP1-3D  &&&&&&&&& 1.11 &&&&&&&&& 1.06 &&&&&&&&& 1.005 \\
SMP2-3D  &&&&&&&&& 1.30 &&&&&&&&& 1.18 &&&&&&&&& 1.010 \\
SMP3-3D  &&&&&&&&& 1.45 &&&&&&&&& 1.30 &&&&&&&&& 1.011 \\
\hline 
\label{Table2}
\end{tabular}
\end{table}

Next, we investigate the average properties of the contact force
network. In Sec.~\ref{Stress-Polydispersity}, we applied the
mean-field approach of Sec.~\ref{Fabric-Polydispersity} to
estimate the isotropic pressure in a given polydisperse granular
sample. However, due to the presence of the normal component of
the contact force $\bar F_n(a)$ in Eq.~(\ref{stress-poly-1}), one
needs to make one further assumption about the particle-size
dependence of $\bar F_n(a)$ to be able to calculate the integral 
and obtain $\langle \widetilde\sigma_{\alpha \alpha} \rangle_{_V}$ 
from the average quantities.

\begin{figure*}[t]
\centering
\includegraphics[scale=0.9,angle=0]{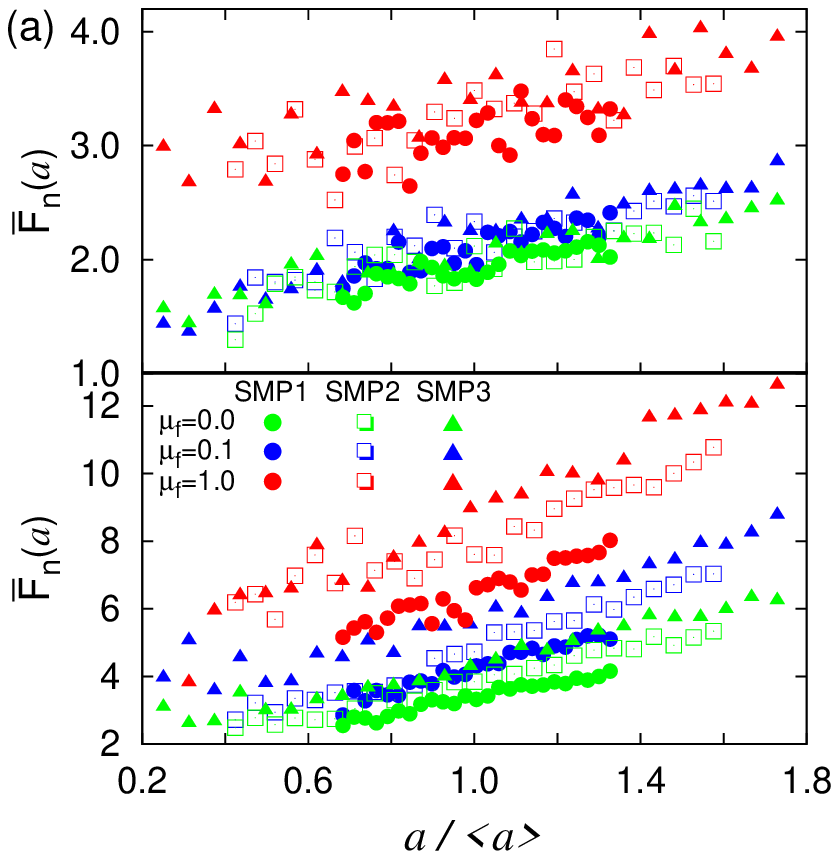}
\includegraphics[scale=0.9,angle=0]{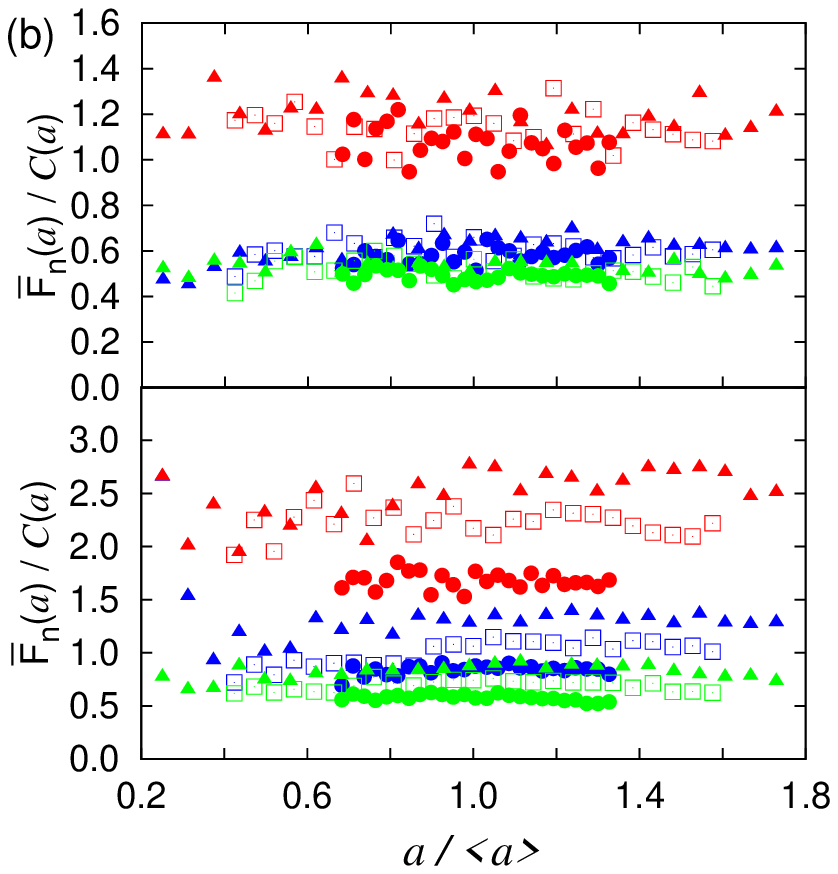}
\caption{(Color online) (a) Normal component of the contact force 
$\bar F_n(a)$, averaged over all particle radii in the same bin, 
in terms of the particle radius $a$ for two-dimensional (top) and 
three-dimensional (bottom) packings constructed with different size 
distribution widths and different friction coefficients $\mu_{_f}$. 
(b) $\bar F_n(a)$ scaled by the contact number $C(a)$ for the 
same set of samples as in (a).}
\label{Fig-NormalForce}
\end{figure*}

\begin{figure}[b]
\centering
\includegraphics[scale=0.65,angle=0]{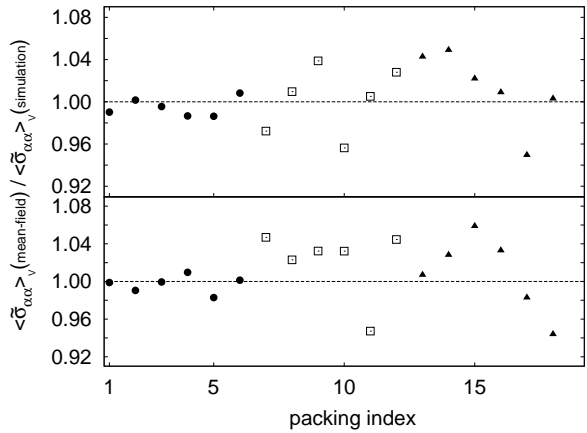}
\caption{The estimated value of the trace of the stress tensor
using Eq.~(\ref{App-10}) (top), and Eq.~(\ref{stress-poly-5}) 
(bottom) divided by $\langle \sigma_{\alpha
\alpha} \rangle_{_V}$ obtained directly from the simulations. 
Each data point corresponds to a different 2D (top) or 3D (bottom) 
packing using symbols as in Table~\ref{Table1}.} 
\label{Fig-Stress}
\end{figure}

\begin{figure}[b]
\centering
\includegraphics[scale=0.65,angle=0]{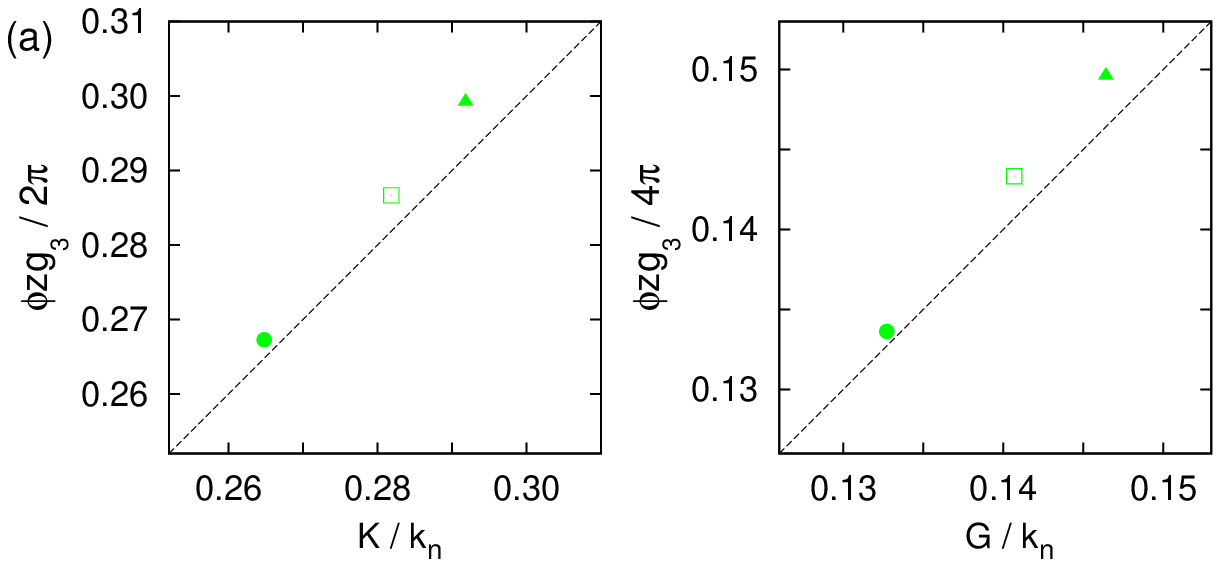}
\includegraphics[scale=0.65,angle=0]{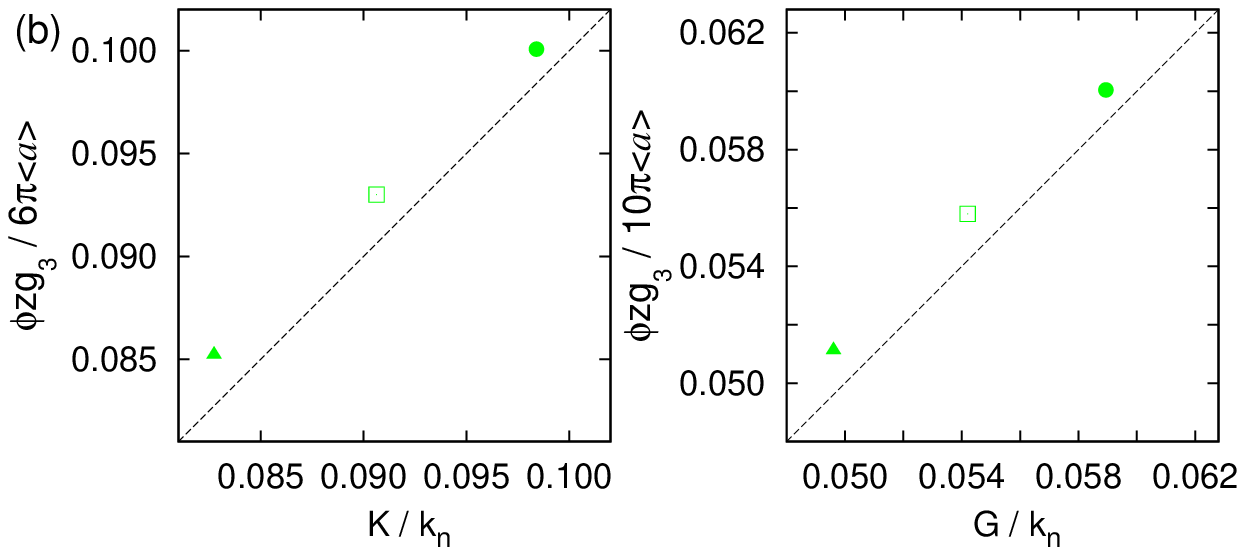}
\caption{(Color online) The estimated values of the bulk $K$ 
(left) and shear $G$ (right) moduli according to Eqs.~(\ref{stiffness-10}) 
and (\ref{stiffness-11}) in 3D [Eqs.~(\ref{App-18}) and 
(\ref{App-19}) in 2D] vs the values obtained from 
the simulation results. Each data point corresponds to one 
frictionless sample and the dashed lines indicate the identity. 
The results are separately shown for (a) 2D and (b) 3D samples. 
The same symbols as in Table~\ref{Table1} are used.}
\label{Fig-StiffnessElements}
\end{figure}

The simulation results [Fig.~\ref{Fig-NormalForce}(a)] reveal
that the average normal force exerted on the particle is an 
increasing function of the particle radius for 2D and 3D 
(the contribution of rattlers is again excluded). With increasing 
friction coefficient and $w$, the average normal force 
increases. This is reminiscent of the behavior of $C(a)$ as a 
function of $a$ (Figs.~\ref{Fig-Coordination-2D} and 
\ref{Fig-Coordination-3D}). Interestingly, the increasing rates 
are similar in both figures. Therefore, it is reasonable to assume 
that the ratio $\bar F_n(a)/C(a)$ is independent of $a$, as already 
observed in 2D \cite{Madadi05}. Figure \ref{Fig-NormalForce}(b) 
confirms the validity of this assumption. We note that the fluctuations 
in Fig.~\ref{Fig-NormalForce}(b) are reduced as the system size 
increases. In Fig.~\ref{Fig-Stress}, we compare the exact 
value of $\langle \sigma_{\alpha \alpha} \rangle_{_V}$ with the 
corresponding value from Eq.~(\ref{stress-poly-5}) [or Eq.~(\ref{App-10})], 
which is obtained based on the above assumption. The results 
are in reasonable agreement with theory for both two- and 
three-dimensional packings, with a standard deviation of $2\%$ to $6\%$ 
for increasing $w$.

Finally, we turn to the calculation of the stiffness tensor 
elements for isotropic materials. We note that to evaluate 
the true elastic moduli, one should apply an incremental strain 
and measure the resulting change of the stress tensor. Alternatively, 
one can read the moduli from the elements of the stiffness tensor, 
assuming the affine motion of the particles, which cannot be taken 
for granted however, and which is the subject of future studies. Here, using 
the packing configuration obtained from the simulation, we calculate 
the elements of the average stiffness tensor via Eq.~(\ref{stiffness-2}). 
Next, the elastic moduli of the packing are calculated using 
Eqs.~(\ref{average-C-elements}), (\ref{Shear-modulus}) and 
(\ref{Bulk-modulus}). The results are then compared to the estimated 
values of the bulk and shear moduli calculated via 
Eqs.~(\ref{stiffness-10}) and (\ref{stiffness-11}) [or 
Eqs.~(\ref{App-18}) and (\ref{App-19})]. 
Figure~\ref{Fig-StiffnessElements} displays the results for several 
two- and three-dimensional packings; the agreement is satisfactory 
within a $5\%$ error (also in the case of frictional packings which 
is not shown here).

According to our analytical results, the ratio between the bulk 
and shear moduli $K/G$ is $5/3$ for isotropic packings independent 
of $z$, $\phi$, and even the size distribution. This suggests 
that in isotropic packings, the ratio between the $P$-wave 
velocity $V_p {=} \sqrt{(K+\frac{4}{3}G)/\rho}$ and the 
$S$-wave velocity $V_s {=} \sqrt{G/\rho}$ is always 
$\sqrt{3}$. An experimental test shows that $V_p/V_s$ for 
a compressed polydisperse packing of glass beads remains 
around $1.7$ over a wide range of pressures from $1$ to 
$7$ MPa \cite{Lebedev11} (see also \cite{Winkler83}). 
Note, however, that anisotropic regular lattice structures 
do not necessarily show the same ratio \cite{Mouraille06}.

\section{Discussion and conclusion}
\label{Conclusion}

In conclusion, a mean-field approach is developed to isolate the 
influence of size polydispersity on the physical properties of 
granular assemblies. We are interested in how the microscale 
quantities are linked to the macroscale ones. 

We find that the trace of fabric and stress tensors factorize into 
the mean packing properties (for example, average coordination 
number, packing fraction, and average normal contact force) and 
dimensionless correction factors, which depend on the moments 
of the particle-size distribution (and approach unity for 
monodisperse packings). The method is extended to estimate the 
elements $C_{_{\!ijkl}}$ of the stiffness tensor. This tensor 
describes the linear affine response of the packing to weak 
external perturbations, when practically the contact network 
between the particles remains unchanged. The elements 
$C_{_{\!ijkl}}$ are also proportional to the average quantities 
and a dimensionless correction factor, which is a function of 
the size distribution. 

\begin{figure}[t]
\centering
\includegraphics[scale=0.68,angle=0]{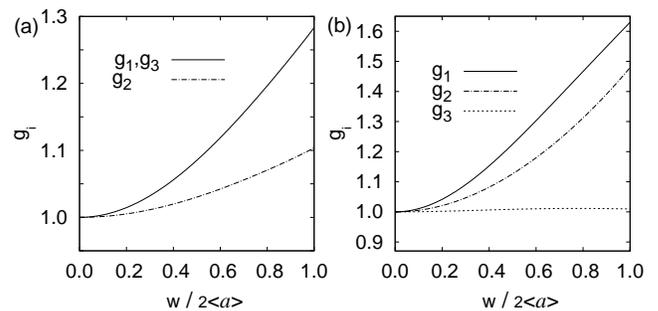}
\caption{The dimensionless correction factors $g_{_i}$ in terms of the 
width $w$ of the uniform size distribution in (a) two and (b) three 
dimensions. $w/2\langle \displaystyle{a}\rangle\!=\!0$ corresponds 
to the monodisperse case.} 
\label{Fig-g-width}
\end{figure}

Numerical simulations illustrate the validity range of our 
analytical predictions and of the assumptions on which the mean-field 
method is based. We note that the deviation of the macroscopic 
quantities of interest from the average packing properties 
increases with increasing the width $w$ of the particle-size 
distribution. Figure~\ref{Fig-g-width} shows the summarized 
correction factors $g_{_i}$ as a function of the width $w$ of 
a uniform size distribution, with the average particle size $\langle 
a \rangle$. Neglecting the correction factors would cause 
remarkable errors, especially for wide distributions. Interestingly, 
$g_{_3}$ is insensitive to the width of the size distribution 
in the 3D case. Therefore, according to Eqs.~(\ref{stiffness-10}) 
and (\ref{stiffness-11}), we expect that the elastic moduli 
of a polydisperse packing of spheres is only moderately 
affected by the choice of $w$. The results of molecular dynamics (MD) 
simulations of soft frictionless spheres imply (see Eq.\ (12) in 
\cite{Goncu10}) that the bulk modulus does not depend on the width of the size 
distribution, in agreement with our analytical results.

The predictive value of this mean-field method should be 
examined also by comparing the theoretical predictions with 
experimental data. For a direct comparison, one needs to 
measure the average packing properties, e.g.\ $z$ and $\phi$, 
which are not easily accessible in experiments (even though 
microcomputed tomography (MicroCT) scan determines the geometry 
with micrometer accuracy nowadays \cite{Aste05}). 
Alternatively, by elimination of $\phi z$ between our 
analytical results, one obtains linear relationships between 
the macroscopic physical properties via some coefficients, 
which depend on the moments of the size distribution. Such 
linear relations between macroscopic quantities have been 
investigated in the literature, e.g.\ between the elastic 
moduli and conductivity \cite{Bristow60} or isotropic 
pressure \cite{Mavko03}, and can be verified experimentally. 
Future studies will more closely examine the nonaffinity of 
deformations of isotropic as well as anisotropic packings 
of frictional and possibly even cohesive particles.

\begin{acknowledgments}
We would like to thank T. Unger, L. Brendel, O. Dur\'an, and M. Lebedev 
for useful discussions and suggestions. M.\;\!M. acknowledges financial 
support by the ANU Digital Core Consortium, and M.\;\!R.\;\!S. and 
D.\;\!E.\;\!W. by DFG Grant No.\ Wo577/8-1 within the priority program 
``Particles in Contact''. S.\;\!L. acknowledges the support of this 
project by the Dutch Technology Foundation STW, which is the applied 
science division of NWO, and by the Stichting voor Fundamenteel 
Onderzoek der Materie (FOM), financially supported by the Nederlandse 
Organisatie voor Wetenschappelijk Onderzoek (NWO).
\end{acknowledgments}

\appendix
\section{Analytical results in two dimensions}
\label{App1}

\begin{figure}
\centering
\raisebox{38ex}{a)}\includegraphics*[width=0.40\textwidth]
{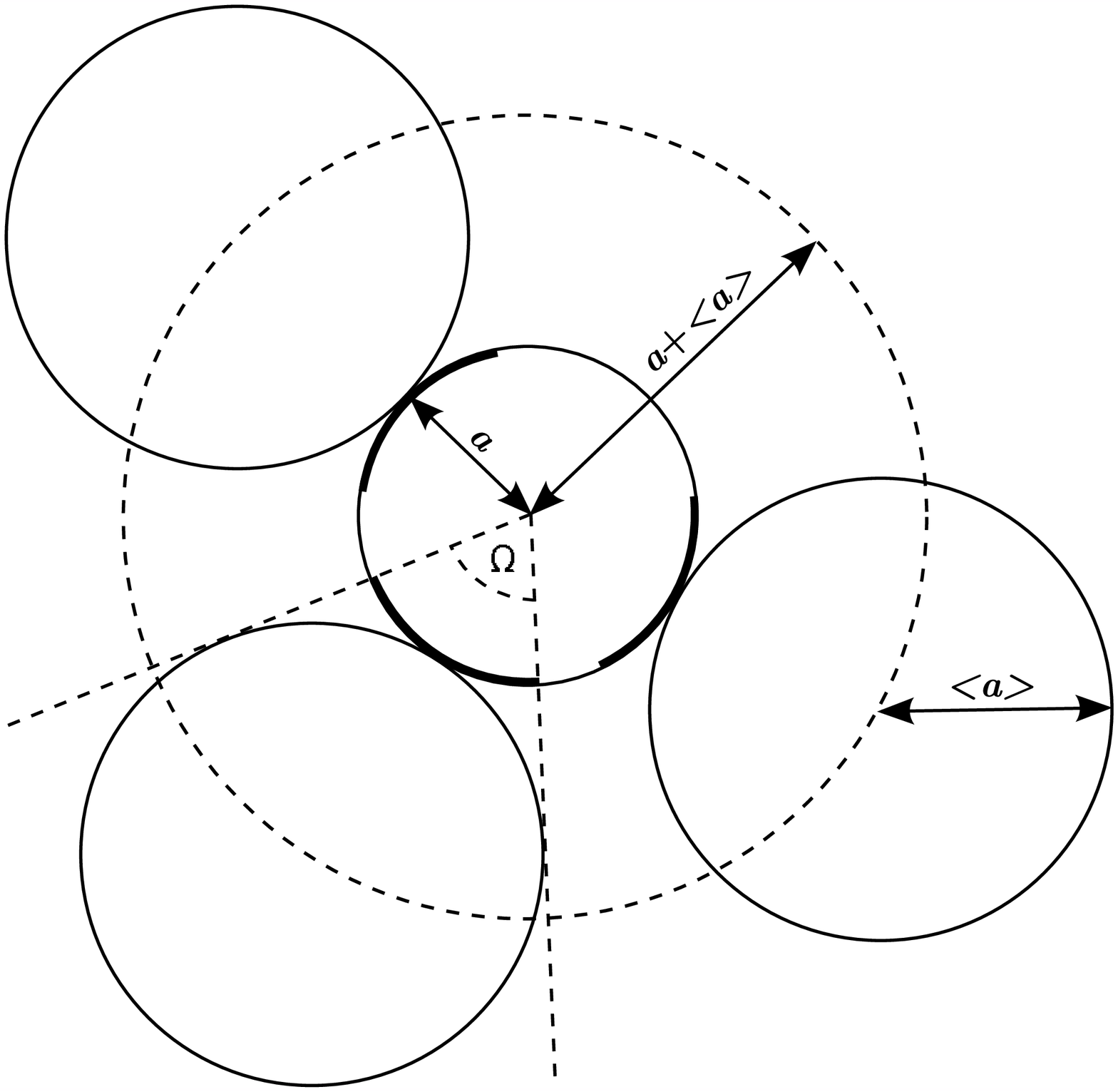}
\raisebox{33ex}{b)}\includegraphics*[width=0.40\textwidth]
{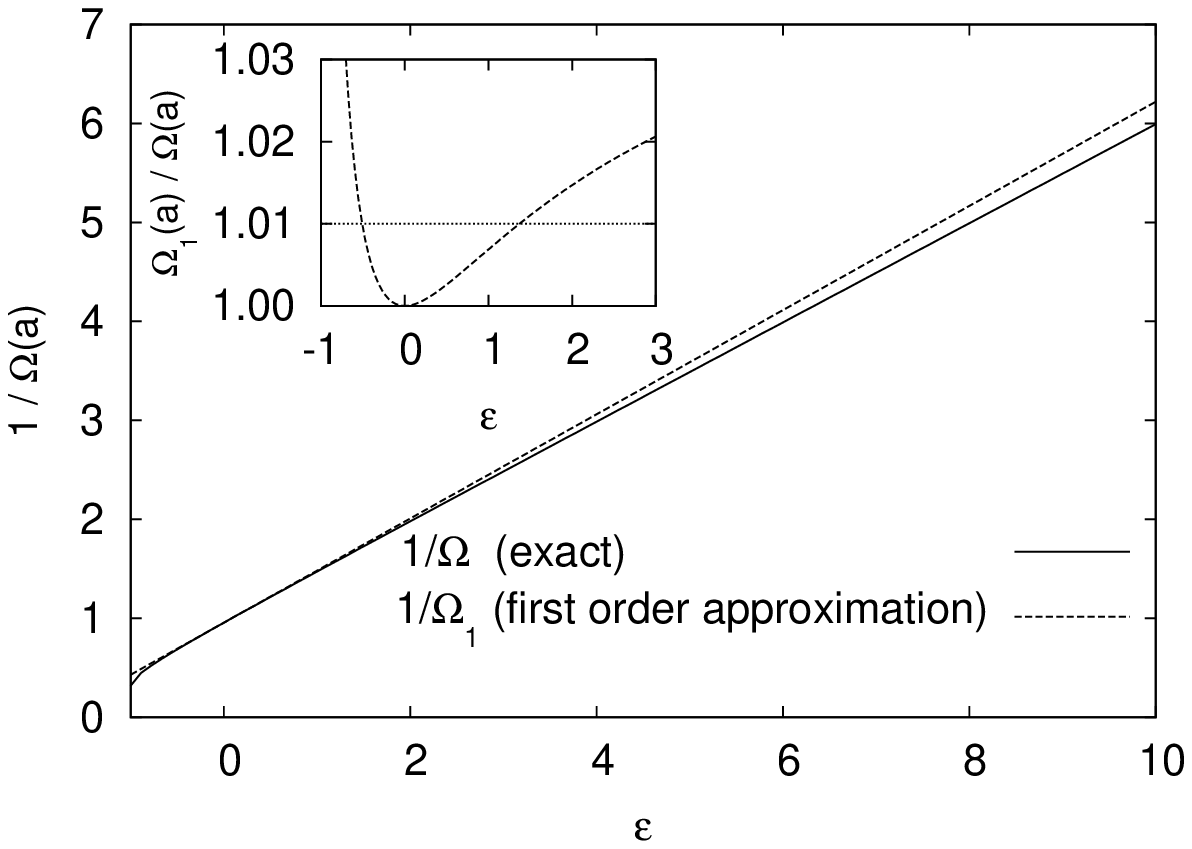}
\caption{(a) A typical particle with radius $a$ surrounded 
by identical particles of average radius $\langle a \rangle$ 
in a 2D packing of disks. The thick solid arcs show the 
shielded surface of the central particle. (b) $1/\Omega(a)$ 
as a function of $\epsilon$ in two dimensions.}
\label{Fig-App1}
\end{figure}

\emph{Fabric tensor .} In a two-dimensional packing of disks 
[Fig.~\ref{Fig-App1}(a)], the surface angle covered by a neighboring 
particle on the reference particle is
\begin{equation}
\Omega(a) = 2 \arcsin \Bigg(\frac{\langle a \rangle}{a+\langle a \rangle} \Bigg),
\label{App-1}
\end{equation}
and the total fraction of the shielded surface is given by
\begin{equation}
c_s(a) = \frac{1}{2 \pi a} \sum_{i=1}^{C(a)}\Omega(a) a = \Omega(a)C(a)/2 \pi.
\label{App-2}
\end{equation}
Assuming that $c_s$ is independent of $a$, one can write the mean 
coordination number $z$ as
\begin{eqnarray}
z = \int_0^\infty \!\!\!\! C(a)f(a)da = 2 \pi c_s q_{_0}.
\label{App-3}
\end{eqnarray}
Equations (\ref{App-2}) and (\ref{App-3}) lead again to Eqs.~(\ref{coordination-number}) 
and (\ref{fabric-meanfield}) for $C(a)$ and $\langle {\mathit h}_{_{\alpha \alpha}} 
\rangle_{_V}$, with the correction factor
\begin{eqnarray}
g_{_1} = \frac{\displaystyle\int_0^\infty V(a) \frac{f(a)}{\Omega(a)}
da}{q_{_0} \displaystyle\int_0^\infty V(a) f(a) da} =
\frac{\langle a^2 \rangle_{_g}}{\langle a^2 \rangle}.
\label{App-4}
\end{eqnarray}
By introducing $\epsilon {=} a / \langle a \rangle{-}1$, we rewrite Eq.~(\ref{App-1}) as
\begin{equation}
\Omega(a) =  2 \arcsin \Bigg(\frac{1}{2+\epsilon} \Bigg),
\label{App-5}
\end{equation}
and approximate $1/\Omega(a)$ to the first order in $\epsilon$ for narrow size 
distributions
\begin{equation}
\frac{1}{\Omega(a)} \simeq A^{\prime}_{_1} + B^{\prime}_{_1}\epsilon,
\label{App-6}
\end{equation}
where $\!A^{\prime}_{_1} \!\!=\! \frac{3}{\pi}$ and $B^{\prime}_{_1} 
\!\!=\! \frac{3\sqrt 3}{\pi^2}$. Figure~\ref{Fig-App1}(b) reveals that 
the approximation has a less than $1\%$ error in the range 
$-0.5\! < \!\epsilon\! < \!1.3$ (or $0.5 \langle a \rangle
\! <\! a \!<\! 2.3 \langle a \rangle$). Hence, $g_{_1}$ for narrow size distributions 
becomes
\begin{equation}
g_{_1} \simeq 1+ \frac{B^{\prime}_{_1}}{A^{\prime}_{_1}}\left(\frac{\big\langle 
\displaystyle{a}^3 \big\rangle}{\big\langle \displaystyle{a} \big\rangle
\big\langle \displaystyle{a}^2 \big\rangle}-1\right).
\label{App-7}
\end{equation}

\emph{Stress tensor .} For a two-dimensional disk, by disregarding the $z$ direction, 
i.e., in the $x\!-\!y$ plane [by requiring $\theta\!=\! \frac{\pi}{2}$ and 
$\bar F_{t_1}^{p}\!\!=\!0$ in Fig.~\ref{Fig-Schematic2}(b)], one obtains
\begin{eqnarray}
&&\hspace{-7.2cm} \widetilde\sigma^p
\!= \frac{a_p}{V_p} \Biggl[ \bar F_n^{p} \sum_{c=1}^{C_p} \!\! \left(
\begin{array}{ccc}
\!\!\! \cos^2(\varphi) & \!\!\!\! \sin(\varphi)\cos(\varphi) \\
\!\!\! \sin(\varphi)\cos(\varphi) & \!\!\!\! \sin^2(\varphi)
\end{array} \!\!\! \right) \!\! \nonumber \\
& + \bar F_{t_2}^{p} \!\!\displaystyle\sum_{c=1}^{C_p}
\!\!\left(
\begin{array}{ccc}
\!\!\!-\sin(\varphi)\cos(\varphi) & \!\!\!\!\cos^2(\varphi) \\
\!\!\!-\sin^2(\varphi) & \!\!\!\!\sin(\varphi)\cos(\varphi)
\end{array}\;\!\!\! \right) \Biggr],
\label{App-8}
\end{eqnarray}
and its trace
\begin{eqnarray}
\widetilde\sigma^{^p}_{_{\alpha \alpha}} = \frac{a_p}{V_p} \sum_{c=1}^{C_p} 
\sum_{\alpha=1}^{D} \biggl( \bar F_n^{p} \; n^{^{pc}}_\alpha 
n^{^{pc}}_\alpha + \bar F_{t_2}^{p} \; n^{^{pc}}_\alpha 
t_{2\alpha}^{^{pc}} \biggr) \nonumber \\
& \hspace{-6.4cm} {=} \displaystyle\frac{a_p}{V_p} \!\! 
\displaystyle\sum_{c=1}^{C_p} \biggl( \!\! \bar F_n^{p} 
|\hat n^{^{pc}}\!|^{^2} {+} \bar F_{t_2}^{p} \hat n^{^{pc}} 
\!\!\!\!\cdot \hat t_{2}^{^{pc}} \!\! \biggr ) {=} \frac{a_p}{V_p} 
\bar F_n^{p} C_p.
\label{App-9}
\end{eqnarray}
Using Eqs.~(\ref{stress-poly-1}) and (\ref{stress-poly-4}), the average 
stress tensor in 2D becomes
\begin{eqnarray}
\langle \widetilde\sigma_{_{\alpha \alpha}} \rangle_{_V} =
\displaystyle\frac{\phi z \; \bar F_n(\langle \displaystyle{a} \rangle) 
\; g_{_2}}{\pi \big\langle \displaystyle{a} \big\rangle},
\label{App-10}
\end{eqnarray}
with
\begin{eqnarray}
g_{_2} = \displaystyle\frac{\pi \langle \displaystyle{a} \rangle 
\displaystyle\int_0^\infty \!\!\! \displaystyle{a} \frac{f(\displaystyle{a})}
{\Omega^2(\displaystyle{a})} \, d\displaystyle{a}}{3q_{_0}\langle 
\displaystyle{a}^2 \rangle}.
\label{App-11}
\end{eqnarray}
By Taylor expansion around $\epsilon \!\!\!=\!\! 0$, we approximate 
$1/\Omega^2(a)$ as
\begin{equation}
\frac{1}{\Omega^2(a)} \simeq A^{\prime}_{_2} + B^{\prime}_{_2}\epsilon,
\label{App-12}
\end{equation}
with $\!A^{\prime}_{_2} \!\!=\! A^{\prime^{^2}}_{_1} \!\!=\! 
\displaystyle\frac{9}{\pi^2}$ and $B^{\prime}_{_2} \!\!=\! 
\displaystyle\frac{18\sqrt 3}{\pi^3}$. Therefore, $g_{_2}$ can be 
approximated by
\begin{equation}
g_{_2} \simeq \frac{B^{\prime}_{_2}}{A^{\prime}_{_2}}+\big(1{-}
\frac{B^{\prime}_{_2}} {A^{\prime}_{_2}}\big) \frac{\big\langle 
\displaystyle{a} \big\rangle^2}{\big\langle \displaystyle{a}^2 
\big\rangle}.
\label{App-13}
\end{equation}

\emph{Stiffness tensor .} Similarly to the three-dimensional analysis 
presented in Sec.~\ref{Stiffness-Tensor}, we approximate the summation 
over neighbors in Eq.~(\ref{stiffness-3}) by $\frac{C(a)}{2\pi}
\int_{_0}^{^{2\pi}} n_{\alpha} n_{\beta} n_{\gamma} n_{\eta} d\theta$, 
which leads to the following reduced stiffness tensor (by mapping $11 
\rightarrow 1$, $22 \rightarrow 2$ and $12 \rightarrow 3$):
\begin{equation}
\langle \mathcal{C} \rangle_{_V} =
\frac{\phi z k_n g_{_3}}{4\pi} \left(
\begin{array}{cccc}
3 & 1 & 0 \\
& 3 & 0 \\
& & 1 \end{array} \right),
\label{App-14}
\end{equation}
with
\begin{equation}
g_{_3} (=g_{_1})= \langle \displaystyle{a}^2 \rangle_{_g}/\langle \displaystyle{a}^2 \rangle,
\label{App-15}
\end{equation}
which for narrow size distributions is approximated as
\begin{equation}
g_{_3} \simeq 1+ \frac{B^{\prime}_{_1}}{A^{\prime}_{_1}}\left(\frac{\big\langle 
\displaystyle{a}^3 \big\rangle}{\big\langle \displaystyle{a} \big\rangle \big\langle 
\displaystyle{a}^2 \big\rangle}-1\right).
\label{App-16}
\end{equation}
In two dimensions, one finds that the Lam\'e constants for frictionless
isotropic packings are
\begin{equation}
\mu = \lambda = (k_n \phi z g_{_3}) \, / \,(4\pi),
\label{App-17}
\end{equation}
and, hence, the shear and bulk moduli are
\begin{equation}
G/k_n = (\phi z g_{_3}) \, / \, (4\pi),
\label{App-18}
\end{equation}
and
\begin{equation}
K/k_n = (\phi z g_{_3}) \, / \, (2\pi).
\label{App-19}
\end{equation}

\end{document}